\documentclass[letterpaper,10pt]{article}
\usepackage{graphicx}
\usepackage{epstopdf, epsfig}
\usepackage{bm}
\usepackage{natbib}
\usepackage{ar}
\usepackage{amsmath}
\usepackage{amssymb}
\usepackage{rotating}
\usepackage{pdflscape}
\usepackage{afterpage}
\usepackage[]{authblk}

\usepackage[left=1in,top=1in,right=1in,bottom=1in]{geometry}

\newcommand{\vect}[1]{\textbf{\textit{#1}}} 
\newcommand{\vort}{\boldsymbol{\omega}} 
\newcommand{\uv}[1]{\hat{\boldsymbol{#1}}} 
\newcommand{\der}[2]{\frac{\mathrm{d}#1}{\mathrm{d}#2}} 
\newcommand{\pder}[2]{\frac{\partial#1}{\partial#2}} 
\newcommand{\Der}[2]{\frac{\mathrm{D}#1}{\mathrm{D}#2}} 
\newcommand{\del}{\boldsymbol{\nabla}} 
\newcommand{\grad}[1]{\del #1} 
\newcommand{\mdiv}[1]{\del\cdot\boldsymbol{#1}} 
\newcommand{\curl}[1]{\del\times\boldsymbol{#1}} 
\newcommand{\md}[1]{\mathrm{d}#1} 
\newcommand{\mD}[1]{\mathrm{D}#1} 
\newcommand{\jb}[1]{[\![#1]\!]} 
\newcommand{\jbb}[1]{\left[\!\!\left[#1\right]\!\!\right]} 
\newcommand{\e}[1]{\mathrm{e}^{#1}} 

\title{The vortex-entrainment sheet in an inviscid fluid: theory and separation at a sharp edge} 

\author{A. C. DeVoria}
\affil{Department of Mechanical and Aerospace Engineering, University of Florida, Gainesville, FL 32611, USA}
\author{K. Mohseni}
\affil{Department of Mechanical and Aerospace Engineering, University of Florida, Gainesville, FL 32611, USA}
\affil{Department of Electrical and Computer Engineering, University of Florida, Gainesville, FL 32611, USA}

\begin{document}

\maketitle

\begin{abstract}
In this paper a model for viscous boundary and shear layers in three-dimensions is introduced and termed a vortex-entrainment sheet. The vorticity in the layer is accounted for by a conventional vortex sheet. The mass and momentum in the layer are represented by a two dimensional surface having its own internal tangential flow. Namely, the sheet has a mass density per-unit-area making it dynamically distinct from the surrounding outer fluid. The mechanism of entrainment is represented by a discontinuity in the normal component of velocity across the sheet. The sheet mass is able to support a pressure jump, which in turn may cause additional entrainment. This feature was confirmed when the model was used to represent the Falkner-Skan boundary layers. The velocity field induced by the vortex-entrianment sheet is given by a generalized Birkhoff-Rott equation with a complex sheet strength. The model was also applied to the case of separation at a sharp edge. There is no requirement for an explicit Kutta condition in the form of a singularity removal as this condition is inherently satisfied through an appropriate balance of normal momentum with the pressure jump across the sheet. A pressure jump at the edge results in the generation of new vorticity. The shedding angle is dictated by the normal impulse of the intrinsic flow inside the bound sheets as they merge to form the free sheet. When there is zero entrainment everywhere the model reduces to the conventional vortex sheet with no mass. Consequently, the pressure jump must be zero and the shedding angle must be tangential so that the sheet simply convects off the wedge face. Lastly, the vortex-entrainment sheet model was demonstrated on two shedding example problems.
\end{abstract}
\section{Introduction}\label{sec:intro}
%
An inviscid fluid is governed by the Euler equations, which permit surfaces of discontinuity or jumps in physical quantities as part of the solution. While these solutions are only mathematical idealizations, their study is of practical use because they can serve as approximations to physical viscous phenomena, particularly at high-Reynolds number ($Re$). For example, combinations of jumps in the fluid density, velocity, pressure, entropy can be cast to represent shocks, boundary/shear layers, or interfaces between fluid media. Here, we focus on the inviscid modeling of viscous layers and flow separation, which has long been a challenging interest in the general fluid dynamics community.

Typically, the high-$Re$ or inviscid approximation $Re\rightarrow\infty$ of viscous layers has been the infinitely thin vortex sheet. Namely, the actual distribution of the velocity and vorticity fields within the layer are lost and the sheet is characterized by a jump in the tangential velocity, which is the local sheet strength representing the integral of vorticity across the layer. In this way, the vortex sheet strength preserves the total circulation. Stated differently, the vortex sheet strength is the circulation per-unit-length of sheet. %

In general, the evolution equation for a conventional vortex sheet is given by a Biot-Savart integral. \cite{Pozrikidis:00a} discusses several of the difficult aspects of computing three-dimensional vortex sheet dynamics including treatments of the principal value integral. For a two-dimensional domain this integral is known as the Birkhoff-Rott equation when expressed in a complex analysis formulation. \cite{Pullin:78a} used this approach and obtained a similarity solution for the shape of a rolled-up semi-infinite free vortex sheet as well as the shapes of vortex sheets shed from the apex of infinite wedges. \cite{Orszag:82a} used a vortex sheet to represent the interface between a stratified flow to study breaking water waves. \cite{JonesMA:03a} computed the truly unsteady vortex sheet shedding from the cusped edges of a moving, finite-chord flat plate where the plate was also represented as a vortex sheet of known geometry. \cite{Mohseni:18l} proposed a model of the local self-induced velocity of a vortex sheet segment that allows integration through the singularity of the Birkhoff-Rott equation. 

Calculation of the dynamics of vortex sheets may be simplified with an approximation obtained by discretizing the vortex sheet into point vortices, a technique that dates back to \cite{RosenheadL:31a} and \cite{WestwaterF:35a} in the 1930s and represents the beginning of what is now referred to as vortex methods \citep{Leonard:80a}. Multi-vortex shedding models have become commonplace \citep[e.g.,][]{ClementsRR:73a,SarpkayaT:75a,KatzJ:81a,LeonardA:93a,Nitsche:94a,MeironD:98a,LlewellynSmith:09a,Mohseni:13ag,EldredgeJD:13a}. These models have been used to great effect for fairly accurate predictions in physical problems, especially the load prediction on and shedding of vorticity from airfoils.

For the conventional vortex sheet the statement of mass conservation is given by the boundary condition of a continuous normal velocity \citep{Saffman:92a}, namely a no-flux condition. Therefore, vortex sheets are mass-less `contact discontinuities' \citep{WuJZ:06a} that remain dynamically indistinct from the surrounding fluid(s). However, real boundary/shear layers contain mass and momentum, which are entrained into the layer. Here, we propose a model that provides a fuller description of the dynamics of a viscous layer by explicitly representing the mass and momentum within the layer. Since entrainment is an inherently viscous process, such a model relaxes the constraint of $Re\rightarrow\infty$ and could be applicable to a range of Reynolds numbers provided the entrainment is modeled sufficiently well.

A physical example that highlights the need for a dynamic model is given by the shedding angle at the \textit{non-cusped} trailing edge of an airfoil. \cite{BasuBC:77a} considered the shedding of a conventional vortex sheet and, based on the works of \cite{GiesingJP:69a} and \cite{MaskellEC:71a}, concluded that the shedding angle must be tangential to one of the edge faces. This result was also proven by \cite{Pullin:78a} for the self-similar shedding problem. 
On the other hand, physical intuition suggests that the shedding angle ought to vary continuously between the tangential limits, which was experimentally observed by \cite{PolingDR:86a}. In this paper we show that if the sheet contains mass then a non-tangential shedding angle is indeed possible.

\cite{Mohseni:17m} recently studied the formation of vortex sheets at the trailing edge of airfoils. They applied conservation laws to a wye-shaped control volume encompassing the two boundary layers at the trailing edge that merge to create the free shear layer. In a similar manner, they also proposed a generalized sheet model for viscous layers, whereby a conventional vortex sheet, with a jump in tangential velocity, can be superimposed with sheets that have jumps in other physical quantities, for example a jump in the stream function representing entrainment into the layer.

The objective of this study is to present and formalize the theory for the surface of discontinuity that we term the vortex-entrainment sheet. In regard to flow separation, the focus in this paper will be on separation at a sharp edge. The paper is organized as follows. In \S\ref{sec:ves} the vortex-entrainment sheet is defined in a three-dimensional domain as a dividing surface containing mass and momentum and thus has its own intrinsic flow within the sheet. \S\S\ref{sec:dyn}-\ref{sec:evo} present the dynamics and kinematics of the sheet as well as the coupling of the internal flow with the flow outside the sheet. The model is then applied as a representation of the self-similar Falkner-Skan boundary layers in \S\ref{sec:FS}. Followed by this are two complementary methods of solving the outer flow via a Laplace equation in \S\ref{sec:Laplace} and a boundary integral formulation in \S\ref{sec:bint}. A general solution algorithm is discussed in \S\ref{sec:algorithm}. The conditions that determine shedding/separation of a sheet from a sharp edge are given in \S\ref{sec:zetao}. Lastly, some example calculations are presented in \S\ref{sec:examps}.

\section{The vortex-entrainment sheet}\label{sec:ves}
Here, we propose the vortex-entrainment sheet as a dynamic inviscid model of viscous layers. To exemplify the physics of entrainment consider the viscous wake left behind a body, where the drag on the body may be determined from a wake survey \citep[][pp. 348-353]{Batchelor:67a}. The expression for the (steady) drag is proportional to the net entrainment rate from infinity, say $Q_\infty$, into the wake: $D=\rho U Q_\infty$. Paired with a similar expression for the lift $L=\rho U\Gamma$, where $\Gamma$ is the circulation, these two conjugate relations have been referred to as the Joukowski-Filon formulae \citep{WuJZ:15a}. There are many explanations of d'Alembert's paradox, but this drag law offers a clear interpretation of resistive forces as due to entrainment of fluid owing to the action of viscosity. 

\begin{figure}
\begin{center}
\begin{minipage}{0.95\linewidth}\begin{center}
\includegraphics[width=0.99\textwidth, angle=0]{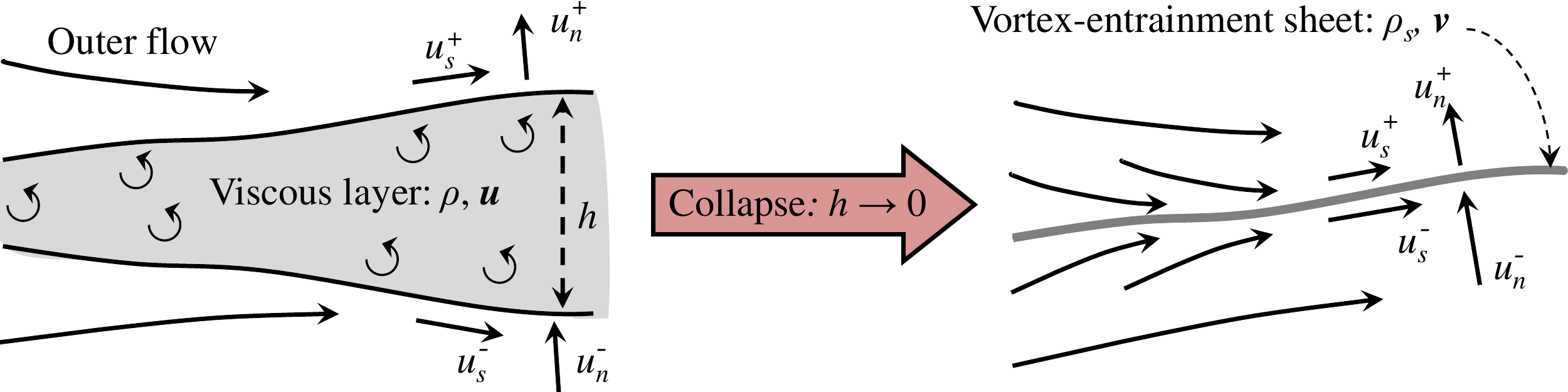}
\end{center}\end{minipage}
\caption{Schematic of the cross-section of a viscous layer of finite thickness $h$ being collapsed to a vortex-entrainment sheet of zero thickness. The fluid inside the viscous layer has density $\rho$ and velocity $\vect{u}$. The mass and momentum in the layer are preserved by assigning to the vortex-entrainment sheet a surface mass density $\rho_s$ and surface velocity $\vect{v}$. The fluid velocities on either side of the sheet, $\vect{u}^\pm$, may be discontinuous.}
\label{fig:collapse}
\end{center}
\end{figure}
Hence, $Q_\infty$ implicitly depends on the Reynolds number, $Re$, through the viscous entrainment process. For example, the wake of a flat plate at zero incidence was calculated by Goldstein via the boundary layer method \citep[][pp. 138-142]{Schlichting:55a}. The total drag on the plate of length $l$ and width $w$ is $D=\rho U(1.33U w\sqrt{\nu l/U})$ from which we identify and rewrite $Q_\infty=1.33w\nu\sqrt{Re_l}$ where $Re_l=Ul/\nu$. The explicit appearance of the viscosity $\nu$ reiterates that entrainment is an inherently viscous process. Hence, the drag could be recovered by equating the net entrainment rate of the vortex-entrainment sheet representing the wake to the value $Q_\infty$. This overly simple example conceptually demonstrates how the vortex-entrainment sheet could be used to provide an improved model of viscous layers. 

As an initial step towards defining the vortex-entrainment sheet we consider the mass within a control volume surrounding a viscous layer and seek to preserve the mass within the layer as its thickness $h$ is collapsed to zero; figure~\ref{fig:collapse}. This concept was applied by \cite{Mohseni:17m} to the case of an airfoil trailing edge that resulted in a discontinuity of the stream function, or equivalently the normal velocity, which represents entrainment into the sheet. The distribution of fluid with density $\rho$ across the layer is preserved by assigning a surface mass density $\rho_s$ with units of mass per-unit-area of sheet. The actual distribution of flow $\vect{u}$ in the layer becomes an `intrinsic' flow $\vect{v}$ confined on the sheet and may possess properties that are different from its bulk surroundings. Surfaces of this type commonly appear in other fields, such as electrodynamics \citep[e.g. a capacitor with surface charge density and discontinuity in the normal component of the electric field,][]{JacksonJD:98a}. While such surfaces have also been considered in fluid mechanics, for example the Boussinesq-Scriven surface fluid model \citep{ScrivenLE:60a}, they are less frequently encountered. 

We will assume that the flow outside the sheet $\vect{u}$ is an irrotational, incompressible flow that may be discontinuous across the sheet. However, we note that this is not to be interpreted as a $Re\rightarrow\infty$ limit, but rather that the viscous, rotational portion of the fluid has been `cut out' and sutured up by the vortex-entrainment sheet. In this regard, it is especially important to note that the vortex-entrainment is \textit{not} a streamline. Next, in \S\S\ref{sec:dyn}-\ref{sec:kin} we more rigorously define the vortex-entrainment sheet.
%

\subsection{Sheet dynamics}\label{sec:dyn}
\begin{figure}
\begin{center}
\begin{minipage}{0.6\linewidth}\begin{center}
\includegraphics[width=0.99\textwidth, angle=0]{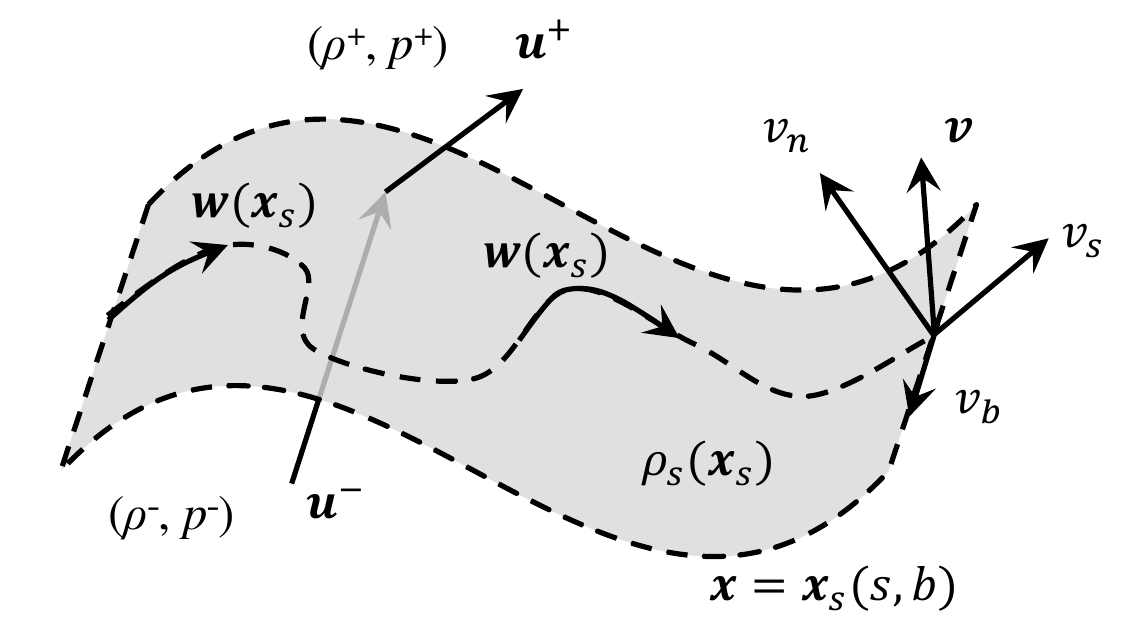}
\end{center}\end{minipage}
\caption{Schematic of a vortex-entrainment sheet (shaded surface) embedded in an outer fluid. The `intrinsic' surface flow $\vect{w}$ is in the tangent space $(s,b)$ of the sheet. The sheet position is given by $\vect{x}=\vect{x}_s(s,b)$. The surface velocity of the sheet is $\vect{v}=\vect{w}+(\vect{v}\cdot\uv{n})\uv{n}$. The sheet has a surface density $\rho_s$ (mass per-unit-area). The outer fluid velocity is $\vect{u}$ and is discontinuous across the sheet. The fluid density $\rho$ and pressure $p$ may also be discontinuous.}
\label{fig:ves}
\end{center}
\end{figure}

Let $\vect{x}$ be the position vector of an arbitrary point in three-dimensional space. The sheet is immersed in an irrotational, incompressible fluid with density $\rho$ and velocity $\vect{u}(\vect{x},t)$, where $t$ is time. A right-handed orthogonal coordinate system ($s$, $n$, $b$) is defined on the sheet such that the basis $\uv{s}$-$\uv{b}$ spans the local tangent manifold that is the sheet and $\uv{n}$ is the corresponding sheet normal vector. The sheet is a two-dimensional surface whose location is specified by $\vect{x}=\vect{x}_s(s,b,t)$; see figure~\ref{fig:ves}.

The jump in a quantity $f$ across the sheet is defined as
\begin{equation}\label{eqn:jump}
\jb{f}=f^{+}-f^{-},
\end{equation}
where $f^{+}$ and $f^{-}$ are the limiting values of $f$ as approached from the different sides of the sheet with $\uv{n}$ pointing to the $(+)$ side. The \textit{surface velocity} on the sheet is $\vect{v}(\vect{x}_s,t)$ and given our choice of coordinates this can be decomposed as
\begin{equation}\label{eqn:tangflow}
\vect{v}=\vect{w}+(\vect{v}\cdot\uv{n})\uv{n}=(v_s\uv{s}+v_b\uv{b})+v_n\uv{n}.
\end{equation}
where $\vect{w}(\vect{x}_s,t)=v_s\uv{s}+v_b\uv{b}$ is the intrinsic velocity of the flow \textit{inside} the sheet and $v_n(\vect{x}_s,t)=\vect{v}\cdot\uv{n}$ is the normal velocity component. The surface operator $\del_s$, which only acts in the tangent space of the sheet, is given by:
\begin{equation}\label{eqn:dels}
\del_s=\left(\frac{1}{h_s}\pder{}{s}\right)\uv{s}+\left(\frac{1}{h_b}\pder{}{b}\right)\uv{b},
\end{equation}
with $h_s$ and $h_b$ as the scale factors for the chosen surface coordinate system. The operator giving the surface material derivative of the quantity $f$ on the sheet is defined as:
\begin{equation}
\Der{_sf}{t}=\pder{f}{t}+\vect{w}\cdot\del_sf.
\end{equation}
Given our specific choice of the coordinate system we have $(v_n\uv{n})\cdot\del_sf=0$ and thus we may write $\vect{w}\cdot\del_sf=\vect{v}\cdot\del_sf$. The reader is referred to \cite{ArisR:62a} and \cite{SlatteryJC:07a} for further details regarding these surface operators as well as the sheet conservation equations discussed next. 

Now, we define the mass within the sheet by integrating across the layer in the normal direction (recall figure~\ref{fig:collapse}) to give
\begin{equation}\label{eqn:rho_s}
\rho_s(\vect{x}_s,t)=\int\rho(\vect{x},t)\md{n}.
\end{equation}
The quantity $\rho_s$ is the sheet mass density and has units of mass per-unit-area of sheet. In this way, the mass that was originally in the viscous layer is preserved by `collapsing' it to the sheet; recall figure~\ref{fig:collapse}. Fluid may enter the sheet via entrainment and the mass conservation equation on the sheet is:
\begin{equation}\label{eqn:mass_3D}
\Der{_s\rho_s}{t}+\rho_s(\del_s\cdot\vect{v})=-\jb{\rho(\vect{u}-\vect{v})\cdot\uv{n}}. 
\end{equation}
The left side has the familiar form of mass conservation in the bulk fluid, but with the understanding of the intrinsic sheet-tangent operators. In contrast, the right side is non-zero with ~$-\jb{\rho(\vect{u}-\vect{v})\cdot\uv{n}}$~ representing a source of $\rho_s$ due to entrainment from the outer flow. While the fluid velocity may have a discontinuity, the sheet velocity is continuous meaning that $\jb{\vect{v}}=0$.

The conservation of momentum on the sheet also takes on a similar form to the usual equation for the bulk fluid, but again with additional source terms associated with the flux and entrainment of momentum from the flow outside the sheet:
\begin{equation}\label{eqn:momentum_3D}
\rho_s\Der{_s\vect{v}}{t}-(\del_s\cdot\mathbf{T}_s+\rho_s\vect{f}_s)=-\jb{\rho(\vect{u}-\vect{v})(\vect{u}-\vect{v})\cdot\uv{n}}-\jb{p}\uv{n}+\jb{\boldsymbol{\tau}},
\end{equation}
where $\jb{p}$ is the pressure jump across the sheet and $\boldsymbol{\tau}=\jb{\mathbf{T}_v\cdot\uv{n}}$ is the shear stress vector with $\mathbf{T}_v$ as the deviatoric stress tensor outside the sheet. In general, the sheet may also have its own phenomena, such as a surface stress tensor $\mathbf{T}_s$, a surface tension force, or a `body' force $\vect{f}_s$ that acts only on the sheet mass density $\rho_s$ \citep{SlatteryJC:07a}. For simplicity we shall neglect each of these terms in (\ref{eqn:momentum_3D}. The surface flow $\vect{v}$ is clearly dynamically coupled to the outer flow $\vect{u}$, which must also be solved. The main unknowns to be solved for are $\vect{u}$, $\rho_s$, $\vect{v}$ and $\jb{p}$. For a free sheet we have $\boldsymbol{\tau}^{+}=\boldsymbol{\tau}^{-}=0$ since the outer flow is potential. When the sheet is bound to a surface its normal velocity will be known from the motion $\vect{U}$ of that surface as $v_n=\uv{n}\cdot\vect{U}$ and the wall shear stress $\boldsymbol{\tau}_w=\jb{\boldsymbol{\tau}}=$ is then appended to the list of unknowns.
 
At this stage it is useful to discuss the occurrence and interpretation of entrainment by examining the jump source term on the right side of (\ref{eqn:mass_3D}). Using the properties of jumps and averages of products \citep[e.g.][p. 34]{WuJZ:06a} we may write
\begin{equation}
\jb{\rho(u_n-v_n)}=\jb{\rho}(\overline{u_n}-v_n)+\overline{\rho}\jb{u_n},
\end{equation}
where $\overline{(\cdot)}=[(\cdot)^{+}+(\cdot)^{-}]/2$ is the arithmetic mean of the values of a quantity across the sheet. Throughout, an overline will indicate this type of average unless otherwise noted. The sheet must coincide with the discontinuity $\jb{u_n}\neq0$ for entrainment to occur meaning that fluid will not pass \textit{through} the sheet. However, since fluid may be entrained \textit{into} the sheet, then $v_n$ will take a value between $u_n^{+}$ and $u_n^{-}$. To elucidate this, consider a weighted average:
\begin{equation}\label{eqn:vn}
v_n=\left[\tfrac{1+\alpha}{2}\right]u_n^{+}+\left[\tfrac{1-\alpha}{2}\right]u_n^{-}=\overline{u_n}+\tfrac{\alpha}{2}\jb{u_n},
\end{equation}
where $\alpha$ is a weighting parameter with $-1\leq\alpha\leq 1$ as similarly used by \cite{Orszag:82a}. Substituting (\ref{eqn:vn}) into the preceding equation gives:
\begin{equation}
\jb{\rho(u_n-v_n)}=\jb{u_n}\left(\overline{\rho}-\tfrac{\alpha}{2}\jb{\rho}\right)=\jb{u_n}\left(\left[\tfrac{1-\alpha}{2}\right]\rho^{+}+\left[\tfrac{1+\alpha}{2}\right]\rho^{-}\right)
\end{equation}
so that fluid with a weighted average density is entrained at a rate $\jb{u_n}$. However, notice the weighting is opposite to that of $v_n$: for example, if $\alpha>0$ then $v_n$ moves with a velocity closer to $u_n^{+}$, but in doing so the sheet entrains or `consumes' more fluid with the density $\rho^{-}$. In a similar fashion, substitution of (\ref{eqn:vn}) into the momentum flux term in (\ref{eqn:momentum_3D}) yields:
\begin{equation}\label{eqn:jump_mom}
\jb{\rho(\vect{u}-\vect{v})(u_n-v_n)}=\jb{u_n}\left(\overline{\rho(\vect{u}-\vect{v})}-\tfrac{\alpha}{2}\jb{\rho(\vect{u}-\vect{v})}\right)
\end{equation}
so that $\jb{u_n}$ simultaneously entrains a weighted average of the difference momentum $\rho(\vect{u}-\vect{v})$ from either side of the sheet. If there is no mass in a free sheet, $\rho_s=0$, then (\ref{eqn:mass_3D}) dictates that $\jb{u_n}=0$. Then (\ref{eqn:momentum_3D}) gives a zero pressure jump $\jb{p}=0$. Of course, these are the kinematical and dynamical features of a conventional vortex sheet. Note that each was consequently obtained from the assumption that $\rho_s=0$.

\subsection{Sheet kinematics}\label{sec:kin}
In order to solve the surface flow equations in (\ref{eqn:mass_3D})-(\ref{eqn:momentum_3D}) we must also determine the outer flow $\vect{u}$. A vector field is determined from its essential characteristics, namely its divergence, curl, discontinuities and boundary values \citep{PhillipsHB:59a}. In general, the field can be expressed by the sum of a divergence-free part and a curl-free part via a Helmholtz-Hodge decomposition, and if the full vector is known on the boundary a harmonic part can be distinguished \citep{NorgardG:13a}. As such, the decomposition can be made unique in several ways depending on the known boundary conditions.

While $\vect{u}$ is an irrotational, incompressible flow with no distribution of vorticity ($\vort$) or dilatation rate ($\Delta$) by definition, it is discontinuous across the sheet. Therefore, the decomposition of $\vect{u}$ has distinct contributions from its discontinuities. These contributions can be rigorously derived \citep[e.g.][]{PhillipsHB:59a}, but for simplicity we employ a straightforward, albeit crude device that yields the same result. If we insist on allowing the fluid to possess a singular curl and divergence we can represent the discontinuity in $\vect{u}$ across the sheet with a Heaviside function. Then the singular parts of the curl and divergence are (see Appendix A):
\begin{eqnarray}\label{eqn:sing_dist}
\vort=\curl{u}=(\uv{n}\times\jb{\vect{u}})\delta(n),\hspace{40pt}
\Delta=\mdiv{u}=(\uv{n}\cdot\jb{\vect{u}})\delta(n),
\end{eqnarray} 
where $\delta(n)$ is the Dirac delta function and $n$ is the sheet normal coordinate. These components of the curl and contribution to the divergence are those affected by normal derivatives across the sheet. The intrinsic surface operators `remove' these singular parts as each is obtained from the appropriate projection of the full spatial operator onto the sheet. These expressions afford a convenient way to define the strengths of the vortex and entrainment sheets as: 
\begin{eqnarray}
\boldsymbol{\gamma}(\vect{x}_s,t) &=& \int\vort(\vect{x}_s,t)\hspace{2pt}\md{n}=\uv{n}\times\jb{\vect{u}}=\jb{u_b}\uv{s}+(-\jb{u_s})\uv{b}, \label{eqn:g_3D} \\
q(\vect{x}_s,t) &=& \int-\Delta(\vect{x}_s,t)\hspace{2pt}\md{n}=-\uv{n}\cdot\jb{\vect{u}}=-\jb{u_n}. \label{eqn:q}
\end{eqnarray}
The negative sign in the definition of $q$ is so that $q>0$ corresponds to entrainment \textit{into} the sheet (also see Appendix B). Relative to the outer flow this will appear as a sink-like motion. The flux into the vortex-entrainment sheet is what sets it apart from a contact discontinuity (no-flux) and from a shock (through-flux). 

The vectorial vortex sheet strength $\boldsymbol{\gamma}=\gamma_s\uv{s}+\gamma_b\uv{b}$ is tangent to the sheet, however, the surface flow may also have finite intrinsic vorticity normal to the sheet: $\omega_n\uv{n}=(\partial v_s/\partial b-\partial v_b/\partial s)\uv{n}$. When (\ref{eqn:sing_dist}) is substituted into the decomposition of $\vect{u}$, the fluid volume integrals reduce to surface integrals over the sheet area $S$. Then using the definitions (\ref{eqn:g_3D})-(\ref{eqn:q}) we can combine the decomposition contributions from the curl/vortex sheet and divergence/entrainment sheet as:
\begin{eqnarray}\label{eqn:BS}
\vect{u}_\omega(\vect{x},t)+\vect{u}_\Delta(\vect{x},t) = \frac{1}{4\pi}\int_S\frac{\boldsymbol{\gamma}(\vect{x}_s,t)\times(\vect{x}-\vect{x}_s)-q(\vect{x}_s,t)(\vect{x}-\vect{x}_s)}{|\vect{x}-\vect{x}_s|^3}\md{A},
\end{eqnarray}
where $\vect{x}_s$ is the position of $S$. The prescribed boundary conditions on a given problem will then determine whether or not these contributions are sufficient to describe $\vect{u}$, for example if a harmonic contribution is needed to impose a boundary condition at infinity.

Finally the circulation vector $\boldsymbol{\Gamma}$ and net entrainment rate $Q$ are defined as:
\begin{eqnarray}\label{eqn:GQ_3D}
\boldsymbol{\Gamma}=\Gamma_s\uv{s}+\Gamma_b\uv{b}=\int\boldsymbol{\gamma}\md{l},\quad\quad
Q=\int_S q\hspace{2pt} \md{A},
\end{eqnarray}
where $l(s,b)$ is a prescribed integration path in the sheet.

\subsection{Evolution equations for the sheet strengths}\label{sec:evo}
Here, the evolution equations for the sheet strengths are given. 
To this end, the jump operator in (\ref{eqn:jump}) is applied to the governing Euler equation for the outer flow $\vect{u}$. \cite{AlbenS:09a} used this process to derive a scalar vortex sheet evolution equation to study two-dimensional flexible propulsive appendages. The jump Euler equation is:
\begin{equation}\label{eqn:Euler}
\pder{\jb{\vect{u}}}{t}=-\jbb{\vect{u}\cdot\grad{\vect{u}}+\grad{\frac{p}{\rho}}}.
\end{equation}
Note that the del operator is inside the jump brackets. While the identity $\vect{u}\cdot\grad{\vect{u}}=\grad{\frac{1}{2}|\vect{u}|^2}$ is valid on each side of the sheet, we cannot use it to write $\jb{\vect{u}\cdot\grad{\vect{u}}}=\grad{\jb{\frac{1}{2}|\vect{u}|^2}}$, since normal derivatives across the sheet are not defined. Therefore, we isolate the tangential and normal components of the right side by making use of the surface-tangent operator $\del_s$ in (\ref{eqn:dels}) to allow the jump operator inside on the tangential part:
\begin{eqnarray}
\jbb{\vect{u}\cdot\grad{\vect{u}}+\grad{\frac{p}{\rho}}}=\del_s\jbb{\frac{1}{2}|\vect{u}|^2+\frac{p}{\rho}}+\jbb{\left(\del\cdot(u_n\vect{u})+\uv{n}\cdot\del\frac{p}{\rho}\right)}\uv{n},
\end{eqnarray}
and we have used the identity $\vect{u}\cdot\del\vect{u}=\del\cdot(\vect{u}\otimes\vect{u})$ for the normal component. We can effectively take the curl and divergence of (\ref{eqn:Euler}) by the cross product and dot product, respectively, with $\uv{n}$. Using the sheet strength definitions in (\ref{eqn:g_3D})-(\ref{eqn:q}) along with the properties of jumps of product quantities and the triple vector product we have the following identities:
\begin{equation}
\jbb{\frac{1}{2}|\vect{u}|^2}=\overline{\vect{u}}\cdot\jb{\vect{u}}=\overline{\vect{u}}\cdot\left(\boldsymbol{\gamma}\times\uv{n}-q\uv{n}\right).
\end{equation}
We then obtain:
\begin{eqnarray}
\uv{n}\times\jbb{\vect{u}\cdot\grad{\vect{u}}+\grad{\frac{p}{\rho}}}&=&\pder{}{b}\left(\overline{\vect{u}}\cdot\jb{\vect{u}}+\jbb{\frac{p}{\rho}}\right)\uv{s}-\pder{}{s}\left(\overline{\vect{u}}\cdot\jb{\vect{u}}+\jbb{\frac{p}{\rho}}\right)\uv{b} \label{eqn:ncross}\\
\uv{n}\cdot\jbb{\vect{u}\cdot\grad{\vect{u}}+\grad{\frac{p}{\rho}}}&=&\pder{}{s}\jb{u_nu_s}+\pder{}{b}\jb{u_nu_b}+\jbb{\pder{}{n}\left(u_n^2+\frac{p}{\rho}\right)}. \label{eqn:ndot}
\end{eqnarray}
In the event that there is no entrainment, then (\ref{eqn:ndot}) would be zero since the pressure jump immediately communicates normal momentum from one side of the vortex-entrainment sheet to the other. To see this, consider the normal component of the sheet momentum equation (\ref{eqn:momentum_3D}):
\begin{equation}\label{eqn:norm_mom}
\rho_s\Der{_sv_n}{t}=-\jb{\rho(u_n-v_n)^2}-\jb{p}=-2\rho (\overline{u_n}-v_n)\jb{u_n}-\jb{p}.
\end{equation}
With $q=-\jb{u_n}=0$ then $\jb{p}$ delivers the acceleration $\mD{_sv_n}/\mD{t}$ to the sheet mass $\rho_s$. On the other hand, when $q\neq0$ then the last term in (\ref{eqn:ndot}) represents a gradient of energy across the sheet meaning that the pressure jump must expend additional energy to also accelerate the \textit{newly} entrained sheet mass. Obviously we cannot evaluate this gradient directly. However, from what has been said about the $q=0$ case we can conclude that the normal component of the jump in the Euler equation is equal to the amount of normal momentum that is lost or entrained into the sheet. Therefore, after defining
\begin{equation}
\mu\equiv\overline{\vect{u}}\cdot\jb{\vect{u}}+\jbb{\frac{p}{\rho}}=\overline{\vect{u}}\cdot\left(\boldsymbol{\gamma}\times\uv{n}-q\uv{n}\right)+\jbb{\frac{p}{\rho}}
\end{equation}
we recombine the tangential and normal components to obtain the evolution equation for the vectorial strength $\boldsymbol{\alpha}=\boldsymbol{\gamma}+q\uv{n}$ of the vortex-entrainment sheet as:
\begin{eqnarray}\label{eqn:evo}
\pder{\boldsymbol{\alpha}}{t}=(\del_s\mu)\times\uv{n}+\left(\jb{\rho(u_n-v_n)^2}\right)\uv{n}.
\end{eqnarray}
Evidently the evolution of the sheets are coupled to each other through the quantity $\mu$, which partly represents the jump in dynamic pressure. For the vortex sheet (\ref{eqn:evo}) represents the familiar result that tangential pressure gradients generate vorticity components bi-normal to the gradient direction \citep{LighthillMJ:63a}. For the entrainment sheet the pressure jump acts to `push' or entrain fluid into the sheet.

\subsection{Physics of inviscid separation: interpretation of the Kutta condition}\label{sec:Kutta}
We now briefly discuss inviscid flow separation at a sharp edge from a physical point of view. Regardless of the shedding problem or its formulation, the requirement of a bounded flow at the sharp edge should be imposed. This constraint is the Kutta condition, which we note has no fundamental basis. However, from (\ref{eqn:ndot})-(\ref{eqn:norm_mom}) we see that the condition of a regular flow is ensured by an appropriate balance of normal momentum with the pressure jump. For the conventional vortex sheet with $\jb{u_n}=\jb{p}=0$ this is automatically satisfied by the condition of mass conservation, namely $\rho_s=0$ .

Next, consider the inviscid mechanism that eliminates the singular pressure \textit{gradient}, but not the jump, and allows the flow to separate from the surface. To see how this neutralization of the normal pressure gradient to its jump value relates to vorticity, we again employ the simple device of allowing the fluid to have a singularity in the form of a delta function as was done for the curl and divergence in (\ref{eqn:sing_dist}). Namely, the singular part of the pressure gradient is $\grad{(p/\rho)}=\uv{n}\jb{p/\rho}\delta(n)$. Since $p$ is \textit{not} $C^2$ at the irregular sharp edge point, then $\del\times\del (p/\rho)$ is not necessarily zero there \citep{GreenbergM:98a} and thus contributes to the (singular) vorticity equation for $\vort=\boldsymbol{\gamma}\delta(n)$ as:
\begin{equation}
\del\times\del \frac{p}{\rho}=\left(\pder{}{b}\jbb{\frac{p}{\rho}}\uv{s}-\pder{}{s}\jbb{\frac{p}{\rho}}\uv{b}\right)\delta(n),
\end{equation}
which are the pressure terms of on the right side of (\ref{eqn:ncross}). Hence, a non-zero pressure jump at the edge \textit{generates new} vorticity and will result in a loading there \citep[e.g. see ][]{SearsWR:56a}. For the conventional vortex sheet $\jb{p}=0$ and while $\vort$ is still singular the above pressure terms are in fact zero. Therefore, in this case the vorticity in a vortex sheet simply convects off the edge.

Lastly, an interesting physical interpretation of separation arises when a wall-bounded vortex-entrainment sheet is represented as a dipole sheet or `double layer' \citep{KelloggO:29a}. The singularity in the pressure gradient field at the edge can be represented by adding the potential of a doublet to the vector potential decomposing $\grad{p}$ \citep{PhillipsHB:59a}. Hence, the finite strength of the doublet would be equal to the pressure jump. Then the neutralization of the pressure gradient singularity `tears apart' the double layer and sheds one sheet into the fluid and the other sheet inside the wall as its image.

\section{Falkner-Skan boundary layers}\label{sec:FS}
Before moving on to the solution of the outer flow in \S\ref{sec:outer}, we apply the vortex-entrainment sheet model to the self-similar Falkner-Skan boundary layer solutions. This also serves the purpose of revealing details of an actual viscous entrainment process. We will obtain the sheet strengths $\gamma$ and $q$ directly from the solutions. The driving outer flow $U(x)=ax^m$ is from left-to-right and the local Reynolds number is $Re_x=Ux/\nu$ and is based on the distance $x$ from the leading edge. 

On the solid surface we have $u=u_s^{-}=0$ and $v=u_n^{-}=0$ due to the no-slip and no-penetration conditions.  The Cartestian components of velocity at the outer edge of the boundary layer are $u^{+}=U(x)$ and $v^{+}=c_vU(x)/\sqrt{Re_x}$ where $c_v$ is a numerical coefficient. In order to calculate the proper vortex and entrainment sheet strengths given in (\ref{eqn:g_3D})-(\ref{eqn:q}), we must use the tangential and normal components. In other words, we must use the known shape of the boundary layer, which we write as $\delta(x)=c_\delta x/\sqrt{Re_x}$. Hence, $u_s^{+}=\vect{u}^{+}\cdot\uv{s}^{+}$ and $u_n^{+}=\vect{u}^{+}\cdot\uv{n}^{+}$ where $\uv{s}^{+}=(\cos\theta,\sin\theta)$ and $\uv{n}^{+}=(-\sin\theta,\cos\theta)$ with $\tan\theta=\md{\delta}/\md{x}$. After some algebra we obtain:
\begin{eqnarray}
\gamma(x)&=&0-u_s^{+}=-U(x)\left[\frac{\sqrt{Re_x}+c_m\frac{c_vc_\delta}{\sqrt{Re_x}}}{\sqrt{(c_\delta c_m)^2+Re_x}}\right] \label{eqn:gBL} \\ 
q(x)&=&0-u_n^{+}=U(x)\left[\frac{c_\delta c_m-c_v}{\sqrt{(c_\delta c_m)^2+Re_x}}\right], \label{eqn:qBL}
\end{eqnarray}
where $c_m=(1-m)/2$. When $Re_x\rightarrow\infty$ the entrainment dies out and the vortex sheet strength becomes $-U(x)$ as expected from the infinite-Reynolds number assumption associated with vortex sheets. As $Re_x$ decreases the entrainment strength becomes relatively more significant. However, the limit $Re_x\rightarrow0$ cannot be interpreted with any real meaning since the boundary layer equations breakdown and are invalid at the leading edge, which leads to the singular behavior of the vortex sheet strength. 

\begin{figure}
\begin{center}
\begin{minipage}{0.49\linewidth}\begin{center}
\includegraphics[width=0.99\textwidth, angle=0]{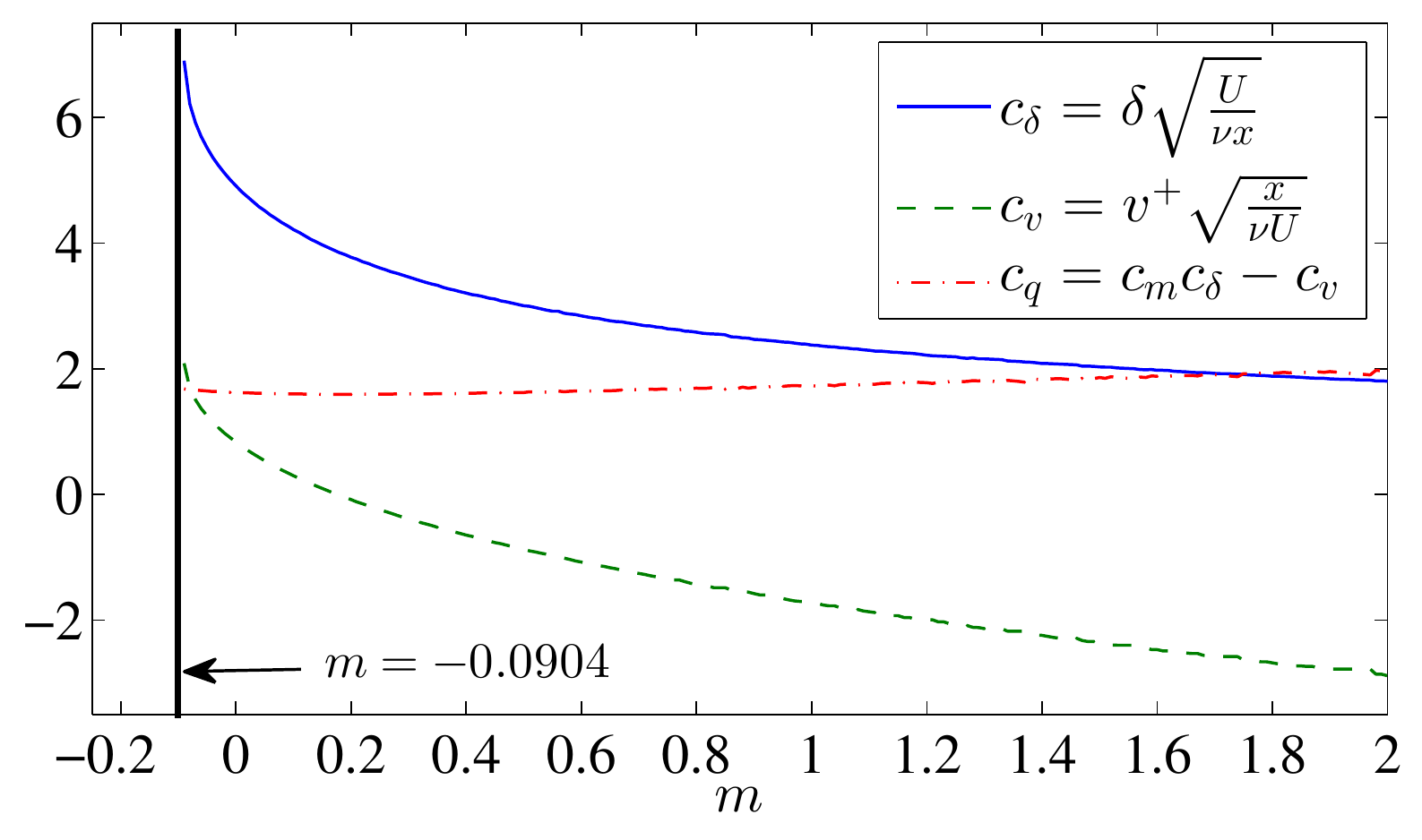}\\ (\textit{a})
\end{center}\end{minipage}
\begin{minipage}{0.49\linewidth}\begin{center}
\includegraphics[width=0.99\textwidth, angle=0]{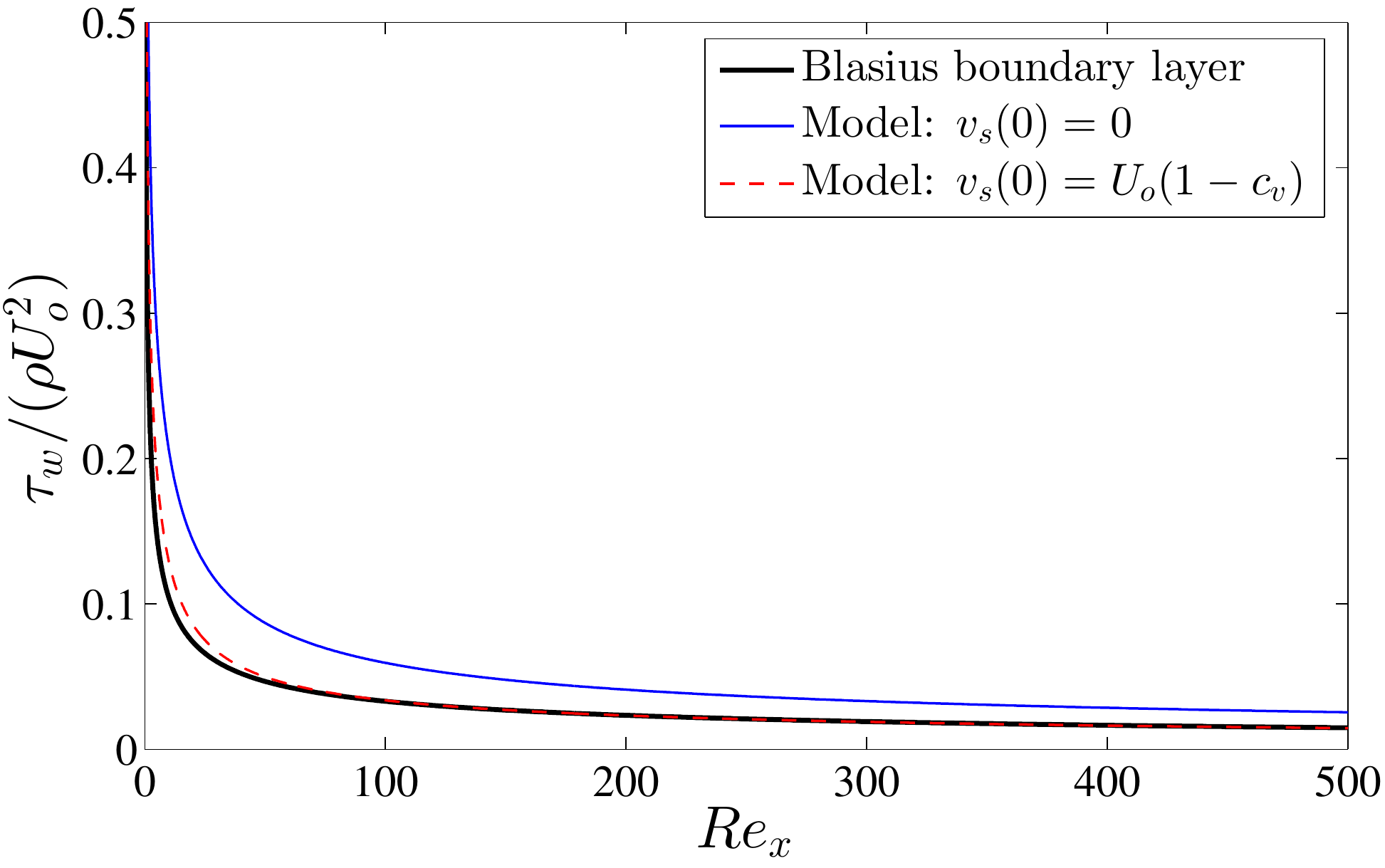}\\ (\textit{b})
\end{center}\end{minipage}
\caption{(\textit{a}) Numerically computed coefficients representing quantities at the edge of the Falkner-Skan boundary layers as functions of $m$ the exponent on the outer flow $U(x)=ax^m$; see text for detail. (\textit{b}) The computed wall shear stress $\tau_w$ predicted by the vortex-entrainment sheet model compared with the Blasius solution: $\tau_{w,B}/(\rho U_o^2)=0.33/\sqrt{Re_x}$. The condition $v_s(0)=U_o(1-c_v)$ accounts for the flux through the leading edge point while $v_s(0)=0$ does not.}
\label{fig:FS_BL}
\end{center}
\end{figure}
Hence, we can safely assume the region of validity to begin at some downstream location such that $\sqrt{Re_x}\gg |c_\delta c_m|$. We computed numerical solutions of the Falkner-Skan equation for $-0.0904\leq m\leq 2$ and indeed find that $|c_\delta c_m|\sim o(10)$; the edge of the boundary layer was defined with the 99\% rule. Using this approximation and upon substituting the definitions of the coefficients $c_\delta$ and $c_v$ the $\sqrt{Re_x}$ factor cancels out and the sheet strengths can be written in terms of the non-dimensional boundary layer thickness $\delta/x$ and vertical velocity $v^{+}/U$ at the edge of the layer as:
\begin{equation}
\gamma(x)\approx -U(x)\left[1+c_m\frac{\delta}{x}\frac{v^{+}}{U}\right],\quad\quad q(x)\approx U(x)\left[c_m\frac{\delta}{x}-\frac{v^{+}}{U}\right].
\end{equation}
The first and second terms in the square brackets of each expression correspond to the horizontal and vertical Cartesian components $u^{+}=U(x)$ and $v^{+}(x)$, respectively. For $q$ there is a competition between the acceleration of the imposed flow `pushing' fluid into the boundary layer and the growth of the layer displacing fluid upwardly. 

Figure~\ref{fig:FS_BL}(\textit{a}) plots $c_\delta$ and $c_v$ against $m$ along with $c_q\equiv c_m c_\delta-c_v$, which is the numerator of (\ref{eqn:qBL}) and represents the sign of $q$. The decreasing behavior of $c_\delta$ is the collapsing boundary layer thickness corresponding to a stronger outer flow and thus larger $Re_x$. For decelerating and slowly accelerating flows $c_v>0$ meaning that displacement outpaces entrainment. There is a positive vertical velocity at the edge of boundary layer and thus an efflux of fluid at infinity. For larger accelerations the trend reverses with $c_v, v^{+}<0$ above $m\approx 0.2$ and there is an influx flow at infinity to compensate for the entrainment. Also, not only is $q>0$ for any $m$, but $c_q\approx 1.74\pm 0.24$ is approximately constant. By (\ref{eqn:gBL})-(\ref{eqn:qBL}) we then have $\gamma\approx -U[1+c_mc_\delta c_v/Re_x]$ and $q\approx 1.74 U/\sqrt{Re_x}$ and both sheets indicate singular behavior near the irregular leading edge point.

Next, we consider the intrinsic flow in the vortex-entrainment sheet representing a boundary layer. For simplicity we take the Blasius boundary layer with $m=0$ and so $U(x)=U_o$. Noting that the problem is two-dimensional, steady and $v_n=0$ since the sheet is bound to the stationary plate ($u_s^{-}=u_n^{-}=0$), then the mass and momentum equations in (\ref{eqn:mass_3D})-(\ref{eqn:momentum_3D}) reduce to:
\begin{equation}
\pder{}{s}(\rho_sv_s)=\rho q,\quad\quad \tau_w=\rho q(\overline{u_s}-v_s)+\rho\gamma\overline{u_n}-\frac{\rho_s}{2}\pder{}{s}\left(v_s^2\right),\quad\quad \jb{p}=2\rho\overline{u_n}q.
\end{equation}
Note that the sheet coordinate is also the plate coordinate $s=x$. The sheet strengths $\gamma$ and $q$ are obtained from (\ref{eqn:gBL})-(\ref{eqn:qBL}) and we can use the plate boundary conditions to give $\overline{u_s}=u_s^{+}/2=-\gamma/2$ and $\overline{u_n}=u_n^{+}/2=-q/2$. Since the entrainment strength decays significantly for $x\gg 1$, the normal momentum equation gives $\jb{p}\approx 0$ which is the boundary layer assumption that the pressure is constant across the layer, or more precisely $O(Re_x^{-1/2})$. Of course this small pressure jump due to entrainment is balanced by an equal and opposite one on the other side of the plate.

For a time-dependent problem, we must be given an initial condition for $\rho_s$ and $v_s$. In this steady case we can obtain the distribution of $\rho_s$ from (\ref{eqn:rho_s}) so that $\rho_s=\rho\delta$. Integrating the mass equation yields $(\rho_sv_s)$ and subsequently we obtain:
\begin{eqnarray}
v_s(x)=\frac{1}{\rho_s}\int_0^x\rho q\md{s}&=&\frac{c_\delta}{\sqrt{Re_x}}\int_0^{Re_x}\rho q\hspace{1pt}\md{(Re_s)}\nonumber \\
&=&\frac{2(c_\delta c_m-c_v)}{c_\delta\sqrt{Re_x}}\left[\sqrt{(c_\delta c_m)^2+Re_x}-c_\delta c_m\right]
\end{eqnarray}
This equation gives $v_s(0)=0$. However, this will violate mass conservation. To see this, recognize that the horizontal velocity $U_o$ is incident to the leading edge point. Although the imposed boundary condition on the flat plate is $v^{-}=0$, the edge of the boundary layer also exists on the plate at $x=0$. Hence, the vertical velocity at the boundary layer edge, $v^{+}(x,\delta(x))=c_vU_o/\sqrt{Re_x}$, also exists at $x=0$. This would indicate that $v^{+}(0,0)$ becomes infinite, but this is not possible. The meaning of this is that limit of $v^{+}(0,0)$ should be a finite, non-zero value and some net flux occurs through the leading edge point given by:
\begin{equation}
v_s(0)=U_o-v^{+}(0,0)
\end{equation}
If we take $v^{+}(0,0)=c_vU_o$ then we obtain $v_s(0)=U_o(1-c_v)$. This means the horizontal velocity $U_o$ is partly `deflected' or displaced into the vertical velocity $c_vU_o$, while the remainder is entrained into the leading edge point. 

Figure~\ref{fig:FS_BL}(\textit{b}) plots the computed wall shear stress $\tau_w$ from the vortex-entrainment sheet model and compares with that from the Blasius solution which is $\tau_{w,B}/(\rho U_o^2)=0.33/\sqrt{Re_x}$. Also shown is the computed $\tau_w$ with the condition $v_s(0)=0$ and we see that there is clearly a larger mismatch. Although, we have accounted for the flux through the leading edge point, there remains the tangential velocity discontinuity of $u^{+}-u^{-}=U_o-0$ so that $\tau_w$ is still singular at this location.

\section{The outer flow soltuion}\label{sec:outer}
This section considers the mathematical formulation for the problem of determining the outer fluid velocity $\vect{u}$. We discuss two different formulations, namely a Laplace equation and a boundary integral formulation. For a given problem, there may be certain advantages to using one or the other or a combination. For simplicity, we will now assume a two-dimensional flow. Hence, the vortex-entrainment sheet becomes a one-dimensional surface with $\uv{b}$ becoming the constant out-of-plane vector $\uv{k}$ so that the fluid and sheet velocities have zero $\uv{b}$ component and those relative to the $\uv{s}$-$\uv{n}$ basis are $\vect{u}=(u_s,u_n)$ and $\vect{v}=(v_s,v_n)$. Likewise $\boldsymbol{\gamma}=\gamma\uv{k}$ and we can deal with the vortex sheet strength as a scalar quantity $\gamma(s,t)=u_s^{-}-u_s^{+}$. The section concludes with a suggested solution algorithm of the dynamically coupled system. 

\subsection{The Laplace equation}\label{sec:Laplace}
The problem is defined by specifying the governing equation and boundary conditions. The conjugacy of the harmonic potential $\phi$ and the stream function $\psi$ allows two equivalent formulations of the problem since both functions satisfy a Laplace equation. Here, we move forward with the $\phi$ framework, but provide a summary showing the duality with the $\psi$ framework. Some useful definitions relating $\Gamma$ and $Q$ to $\gamma$ and $q$ as well as to $\phi$ and $\psi$ may be found in Appendix B.

\begin{figure}
\begin{center}
\begin{minipage}{0.38\linewidth}\begin{center}
\includegraphics[width=0.99\textwidth, angle=0]{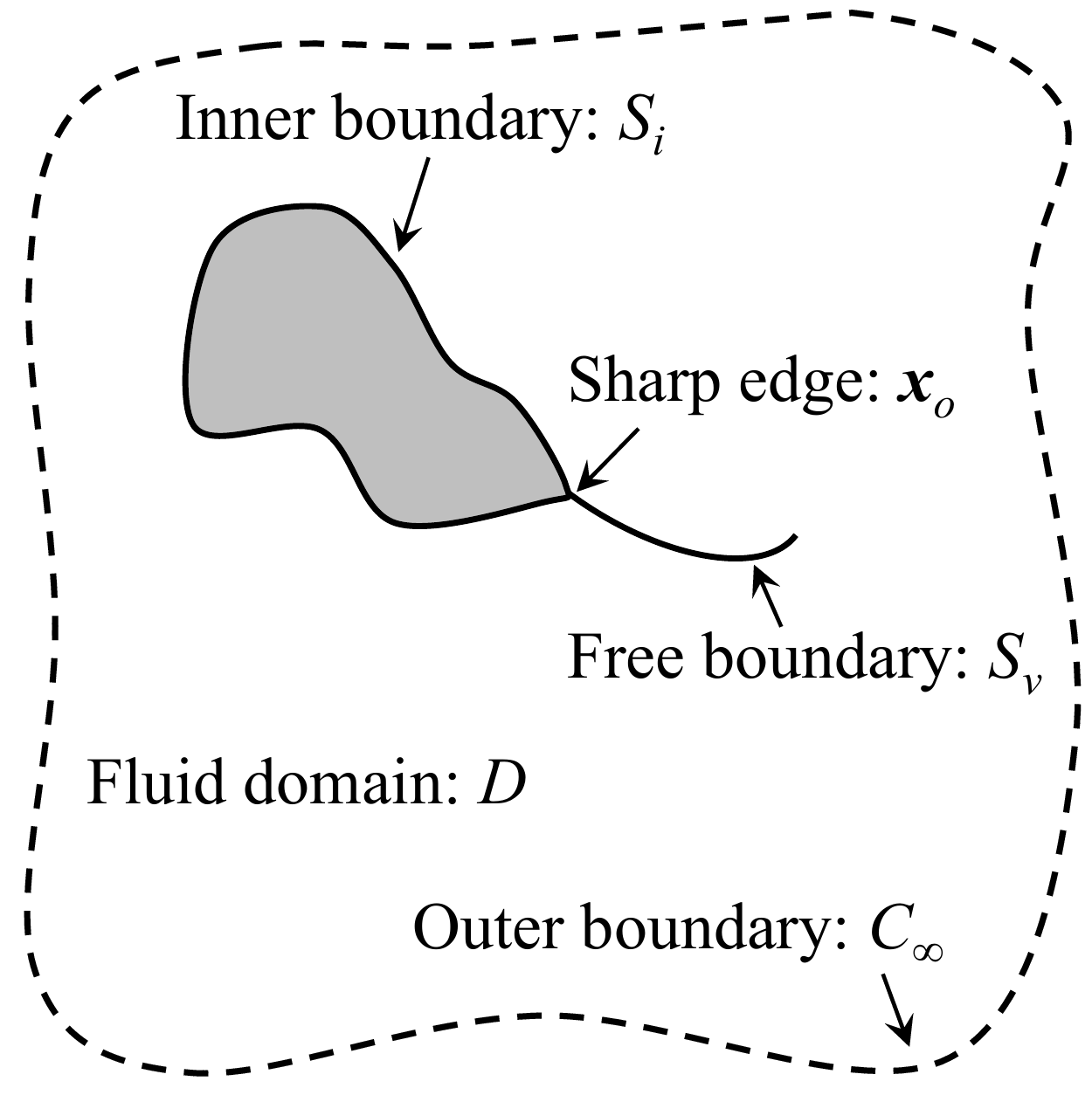}\\ (\textit{a})
\end{center}\end{minipage}
\hspace{50pt}
\begin{minipage}{0.38\linewidth}\begin{center}
\includegraphics[width=0.99\textwidth, angle=0]{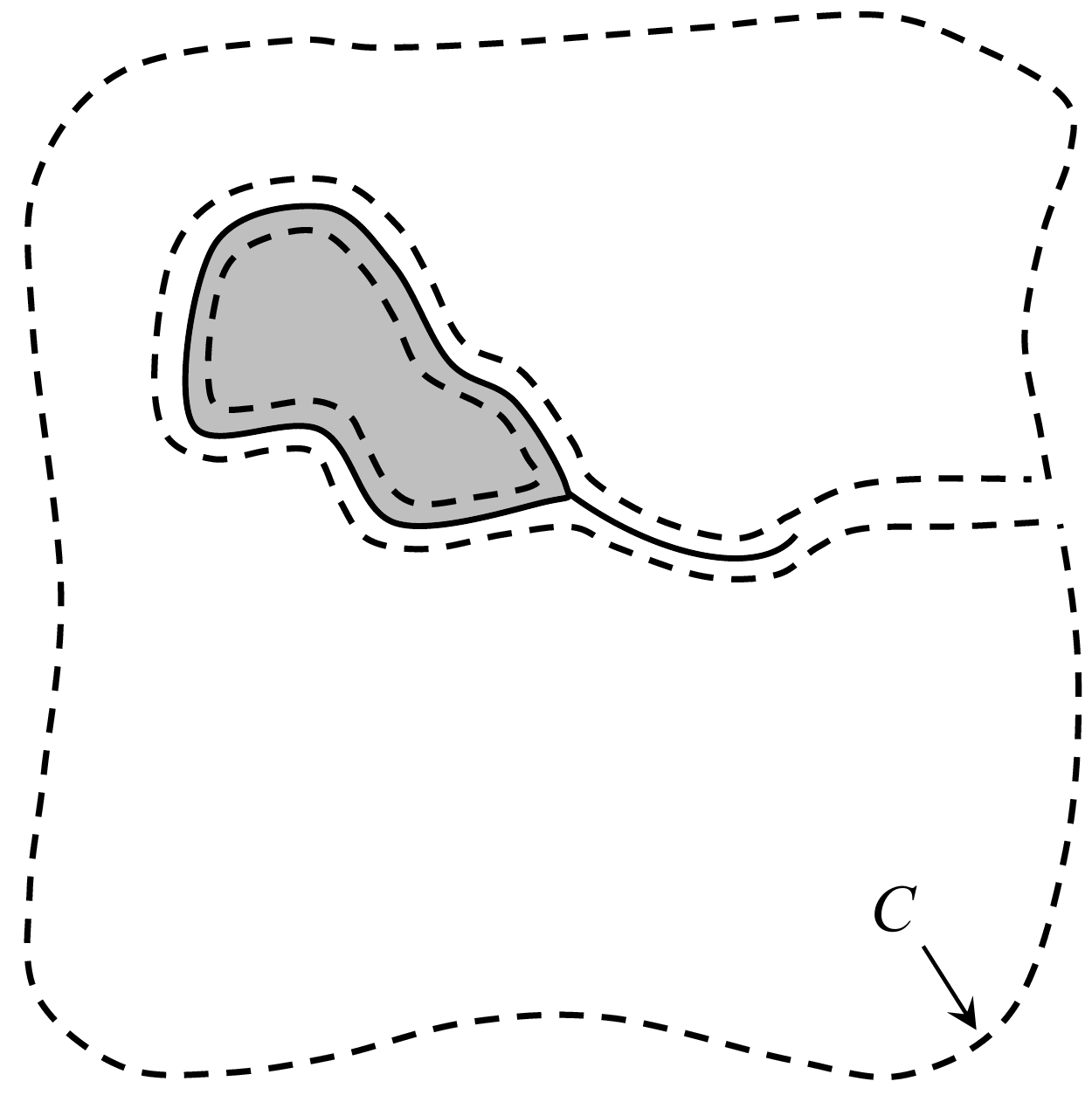}\\ (\textit{b})
\end{center}\end{minipage}
\caption{(\textit{a}) Definition of the fluid domain $D$ and its boundary $\partial D =S_i\cup S_v\cup C_\infty$. $S_i$: inner boundary, $S_v$: free sheet boundary, $C_\infty$: outer boundary, $\vect{x}_o$: sharp edge location. $C_\infty$ is arbitrary and is not a surface of discontinuity. (\textit{b}) The contour integral path $C$ laying entirely in the analytic fluid domain. The path is able to cross itself at the irregular point $\vect{x}_o$ to pass on both sides of $S_i$.}
\label{fig:infinity}
\end{center}
\end{figure}
Let the \textit{open} fluid domain be $D$ with boundary/closure defined as the union $\partial D=S_i\cup S_v\cup C_\infty$ (figure~\ref{fig:infinity}\textit{a}), where $S_i$ and $S_v$ correspond to the vortex-entrainment sheets representing the surface with a sharp edge and the free sheet shed from the edge, respectively, and $C_\infty$ is a large arbitrary boundary `at infinity' that does not represent a surface of discontinuity. Usually, the boundary condition on $S_i$ and $S_v$ is prescribed as continuity of normal velocity, namely $(\partial\phi/\partial n)^{+}=(\partial\phi/\partial n)^{-}$. Under this condition no entrainment occurs, and for a solid surface the normal velocity of that surface, $U_n=(\partial\phi/\partial n)^{\pm}$, is specified. 

However, the vortex-entrainment sheet is characterized by jumps in both the tangential and normal velocity components across the sheet. Hence, the values of $(\partial\phi/\partial n)^{\pm}$ and thus the entrainment rate $q$ are assumed as given boundary conditions (however, see end of \S\ref{sec:algorithm}) . The special conditions for the sharp edge will be discussed in \S\ref{sec:zetao}. With a fluid of infinite extent, the specification of the flow there serves as a boundary condition on $C_\infty$, which may be expressed using an asymptotic form of the potential \citep[][chap. 2.10]{Batchelor:67a}. 
The statement of the problem thus far is:
\begin{align}
\nabla^2\phi=0\quad\quad &\vect{x}\in D \label{eqn:PDE}\\
\phi(\vect{x})-\phi_\infty(\vect{x})\rightarrow C_\phi+\frac{\Gamma_\infty\theta}{2\pi}-\frac{Q_\infty}{2\pi}\log(r)+O(r^{-1})\quad\quad &\vect{x}\in C_\infty \label{eqn:BCinf}\\
\left.\pder{\phi}{n}\right|^{\pm}=u_n^{\pm}(\vect{x})\quad\quad &\vect{x}\in S_i\cup S_v \label{eqn:BCv}
\end{align}
where $C_\phi$ is a constant, $\phi_\infty$ is an external potential representing a non-zero flow at infinity, $\Gamma_\infty$ and $Q_\infty$ are, respectively, the net circulation along $C_\infty$ and the net flux across $C_\infty$ \textit{into} $D$. In the case of a semi-infinite geometry $S_i$ intersects $C_\infty$ at infinity where the potential asymptotically becomes compatible with (\ref{eqn:BCinf}). 

Now consider a contour $C$ residing entirely within the analytic fluid region and bounding the open domain $D$ as defined in figure~\ref{fig:infinity}(\textit{b}). By mass conservation the total flux $Q$ across $C$ must be zero. This yields a `compatibility condition' required for a solution $\phi$ to this Neumann problem to exist \citep[e.g.][]{StakgoldI:68a}. Similarly, by circulation conservation (Kelvin's theorem) the total circulation $\Gamma$ around $C$ must also be zero. Since $C=\partial D$, then with $\vect{u}=\grad{\phi}$ these conditions can be written as:
\begin{equation}\label{eqn:C}
Q=-\oint_C\vect{u}\cdot\uv{n}\md{s}=-\oint_C\pder{\phi}{n}\md{s}=0,\quad\quad\quad \Gamma=\oint_C\vect{u}\cdot\uv{s}\md{s}=\oint_C\pder{\phi}{s}\md{s}=0
\end{equation}
where $s$-$n$ is an orthogonal coordinate system along $C$. The $Q$ equation can be derived from the Laplace equation itself upon using the divergence theorem. In general, the integration path is allowed to cross itself at the irregular point $\vect{x}_o$ in order to trace both sides of $S_i$. When $S_i$ represents a solid, no-slip surface this is not necessary because $\grad{\phi}=\vect{U}$ on the inside of $S_i$. Since $\partial\phi/\partial n$ is given everywhere on the boundary, then $\vect{u}(\vect{x})=\grad{\phi}$ is uniquely determined throughout the domain $D$.

\begin{table}\label{tab:phi_psi}
\begin{center}
\begin{tabular}{lll}
\vspace{5pt}
Conjugate function: \quad &\quad $\phi$ \quad &\quad $\psi$ \\ \vspace{5pt}
Preserved quantity: \quad &\quad mass flux: $\mdiv{u}=0$ \quad &\quad circulation: $\curl{u}=0$ \\ \vspace{5pt}
Governing equation: \quad &\quad $\grad{^2\phi}=0$ \quad &\quad $\grad{^2\psi}=0$ \\ \vspace{5pt}
Boundary condition: \quad &\quad $\pder{\phi}{n}=f(s)$ \quad &\quad $\pder{\psi}{n}=g(s)$ \\ \vspace{5pt}
Velocity expression: \quad &\quad $\vect{u}=\grad{\phi}$ \quad &\quad $\vect{u}=\del\times(\psi\uv{k})$ \\ \vspace{5pt}
Compatibility condition: \quad &\quad $Q=-\oint\pder{\phi}{n}\md{s}=0$ \quad &\quad $\Gamma=\oint\pder{\psi}{n}\md{s}=0$ 
\end{tabular}
\caption{Conjugate frameworks for potential flow governed by Laplace equations for the harmonic potential $\phi$ and the stream function $\psi$.}
\end{center}
\end{table} 
Lastly, in the $\psi$ problem formulation the normal derivative $\partial\psi/\partial n$ is given everywhere on the boundary and corresponds to the tangential velocity component. The conditions in (\ref{eqn:C}) can be written with derivatives of $\psi$ via the Cauchy-Riemann relations. Table~1 summarizes the two different frameworks, which are essentially two different, yet unique decompositions \citep{NorgardG:13a}.

\subsection{Boundary integral formulation}\label{sec:bint}
Now we write the outer flow $\vect{u}$ as a complex conjugate velocity field $w=u-iv$ and the contribution from the vortex-etnrainment sheet is obtained from a Cauchy-type boundary integral. 
\cite{JonesMA:03a} obtained an elegant solution in this manner for the case of a moving flat plate with no entrainment by representing both the free and plate-bound vorticity as conventional vortex sheets. Namely, the two vortex sheets on each side of the plate were combined into a single sheet by coupling the plate and fluid velocities via a tangential boundary condition representing no-slip. 

First, the velocity induced by the vortex-entrainment sheet in the two-dimensional version of (\ref{eqn:BS}) is expressed in a convenient complex form. Namely, $\vect{x}$ is replaced with $z=x+iy$ and the sheet position $\vect{x}_s$ with $\zeta=x_s(s)+iy_s(s)$ where $s$ is the arclength coordinate. Introducing $\theta(\zeta)$ as the angle measured from the horizontal to the local unit tangent vector on the sheet we have $\partial\zeta/\partial s=\e{i\theta(s)}$. The conjugate velocity $w(z)=u-iv$ at an analytic fluid point can then be written as:
\begin{eqnarray}\label{eqn:BRE}
w(z)-w_\infty(z)=\frac{1}{2\pi i}\int_S\frac{\gamma(s)-iq(s)}{z-\zeta(s)}\md{s}=\frac{1}{2\pi i}\int_S\frac{\chi(\zeta)}{z-\zeta}\md{\zeta},
\end{eqnarray}
where $\chi(\zeta)=(\gamma-iq)\e{-i\theta}$ is the complex sheet strength and the integration is over all sheets: $S=S_i\cup S_v$. Also, $w_\infty(z)=\md{\Phi_\infty}/\md{z}$ is the velocity of an external flow that may not decay at infinity. This singular integral is a generalized Birkhoff-Rott equation for the vortex-entrainment sheet with complex strength. Following \cite{MuskhelishviliNI:46a} the left and right sides of the sheet are relative to an observer traversing the sheet in the positive direction of integration in (\ref{eqn:BRE}) and the limits of any quantity approached from the left and right will be denoted with ($+$) and ($-$) superscripts, respectively. For any point on a sheet besides the sharp edge at $\zeta_o$ we have the following by the smooth-arc Plemelj forumlae (see comment in Appendix C):
\begin{equation}\label{eqn:Plemelj}
\left.\begin{array}{ll}\chi(\zeta)=(w^{-}-w^{+})=(\gamma-iq)\e{-i\theta}\\
\\
w(\zeta)=\tfrac{1}{2}(w^{+}+w^{-})=(\overline{u_s}-i\overline{u_n})\e{-i\theta} \end{array}\right\}\quad\text{for}\quad\zeta\in S_i\cup S_v\backslash\zeta_o,
\end{equation}
where $\overline{u_s}$ and $\overline{u_n}$ are the averages of the fluid velocity components. With this formulation it is easy to see how the sheet strength $\chi(\zeta)$ relates to the jump in the complex potential $\Phi=\phi+i\psi$ across the sheet (see Appendix B):
\begin{eqnarray}
\Phi^{-}-\Phi^{+}&=&\left(\phi^{-}-\phi^{+}\right)+i\left(\psi^{-}-\psi^{+}\right)=\Gamma-iQ \\
\pder{}{s}\left(\Phi^{-}-\Phi^{+}\right)&=&\left(u_s^{-}-u_s^{+}\right)-i\left(u_n^{-}-u_n^{+}\right)=\gamma-iq. %
\end{eqnarray}
Note that $\partial\jb{\Phi}/\partial n=i(\partial\jb{\Phi}/\partial s)$, again highlighting the conjugacy of the problem. 

Now, the normal boundary condition (\ref{eqn:BCv}) on a sheet $S$ can be expressed by:
\begin{equation}
\text{Re}\left\{\tfrac{1}{2}(w^{+}+w^{-})i\e{i\theta}\right\}=\tfrac{1}{2}(u_n^{+}+u_n^{-})=\overline{u_n}
\end{equation}
where each of the functions is evaluated at a position $\zeta$ on the sheet. Since we have supposed the normal velocities $u_n^{\pm}$ to be given, and thus the entrainment strength $q$ to be known, the boundary condition can be rearranged as:
\begin{eqnarray}\label{eqn:norm_BC}
f(\zeta)\equiv \text{Re}\left\{\frac{\e{i\theta(\zeta)}}{2\pi}\int_{S}\frac{\gamma(\xi)\e{-i\theta(\xi)}\md{\xi}}{\zeta-\xi}\right\} = &&\overline{u_n}(\zeta)-w_{\infty,n}(\zeta) \nonumber \\
&+&\text{Re}\left\{\frac{\e{i\theta(\zeta)}}{2\pi}\int_{S}\frac{iq(\xi)\e{-i\theta(\xi)}\md{\xi}}{\zeta-\xi} \right\}
\end{eqnarray}
where $\xi\in\mathbb{C}$ is a dummy integration variable along the sheet. This equation is to be solved for $\gamma(\zeta)=\gamma(s)$. The inversion formula and subsequent manipulations used by \cite{JonesMA:03a} are specific to the case of a plane boundary with $\theta(\zeta)$ constant along the sheet. As such, (\ref{eqn:norm_BC}) is more amenable to numerical solution via, for example, expansion of $f(\zeta)$ as a Chebyshev series, which was also done by Jones as well as others since the series converges rapidly for smooth functions \citep{AlbenS:08a}.

\subsection{Solution algorithm}\label{sec:algorithm}
We now present a brief solution algorithm for the coupled system of the vortex-entrainment sheet and the outer fluid. First, the outer flow $\vect{u}$ is solved according to the specified normal boundary conditions with either the Laplace formulation (\ref{eqn:PDE})-(\ref{eqn:BCv}) or the boundary integral formulation (\ref{eqn:BRE})-(\ref{eqn:norm_BC}). For a free sheet we assume that it acquires circulation only as a result of having been shed from a body. Hence, $\gamma$ and $q$ will be known for these sheets. For a surface-bound sheet, the solution of the Laplace formulation will yield the velocity on the fluid side of the sheet, say $\vect{u}^{+}$. Then, the additional boundary conditions on the surface side of the sheet are imposed to yield the sheet strengths. More specifically, we have $\vect{u}^{-}=\vect{U}$ with $\vect{U}$ as the known surface velocity and thus $\vect{u}^{-}-\vect{u}^{+}=\gamma\uv{s}+q\uv{n}$. The imposed conditions will typically be the no-slip and no-penetration conditions for a solid body, but in general we may model actual fluid slip on or porosity of the surface. 

When the boundary integral formulation is used the specification of $u_n^{-}=U_n$ on the surface side of the sheet will uniquely solve the flow inside that surface. Then the solution for $\gamma$ will correspond to the $u_s^{-}-u_s^{+}$ of these two flows. To obtain the desired vortex sheet strength, the tangential boundary condition can be imposed by using the relative velocity $(\overline{u_s}-U_s)$ in the evolution equation (\ref{eqn:evo}) as was done by \cite{JonesMA:03a}. For a solid body, the harmonic potential $\phi_b$ inside the body will be such that $\grad{\phi_b}=\vect{U}$; this is most easily seen in the reference frame of the body or for a stationary body where $\phi_b=const.$ and thus the flow inside vanishes \citep{Lamb:45a}.

Equipped with $\vect{u}$, $\gamma$ and $q$ the solution for the sheet flow quantities can be obtained. The sheet mass and momentum equations (\ref{eqn:mass_3D})-(\ref{eqn:momentum_3D}) along with the evolution equation (\ref{eqn:evo}) represent 4 equations for the 4 unknowns $\rho_s$, $\vect{v}=(v_s,v_n)$ and $\jb{p}$; in three dimensions the system is augmented by the third component $v_b$ and the corresponding vortex sheet strength component $\gamma_s$. Again, these equations are coupled through the pressure jump and the entrainment strength. An initial condition at $t_0$ is required for the intrinsic flow quantities $\rho_s$ and $\vect{v}=\vect{w}+v_n\uv{n}$. In most cases, such as a flow started from rest, these initial conditions will simply be $\rho_s(\vect{x}_s,t_0)=\vect{w}(\vect{x}_s,t_0)=0$ since no entrainment will have occurred yet anywhere in the domain. 

Lastly, we note that it is theoretically possible to obtain both the vortex and entrainment sheet strengths if supplemental conditions are given elsewhere in the domain. In particular, if we are given the full velocity vector on $C_\infty$ `at infinity,' recall figure~\ref{fig:infinity}(\textit{a}), then we may remove the requirement of \textit{a priori} specification of the normal boundary conditions on the sheets. By (\ref{eqn:BCinf}) and (\ref{eqn:BRE}) we obtain:
\begin{equation}\label{eqn:inv}
w(z)-w_\infty(z)=\left(\frac{\Gamma_\infty-iQ_\infty}{2\pi i R}\right)\e{-i\theta(z)}=\frac{1}{2\pi i}\int_S\frac{\gamma(s)-iq(s)}{z-\zeta(s)}\md{s}
\end{equation}
for $|z|=R\rightarrow\infty$. The above integral is non-singular since $z\notin S$ and we note that by writing $z-\zeta(s)\approx R\e{i\theta(z)}$ then:
\begin{equation}
\Gamma_\infty-iQ_\infty=\int_S(\gamma(s)-iq(s))\md{s},
\end{equation}
which states that the circulation and flux in the sheets is equal to that at infinity or that the total of these quantities in the domain $D$ is zero; recall (\ref{eqn:C}). This condition makes the inversion formula of the \textit{singular} integral equation in (\ref{eqn:norm_BC}) unique \citep{JonesMA:03a}. The boundary conditions at infinity, for example by specification of $\Gamma_\infty$ and  $Q_\infty$ on $C_\infty$ would then determine the sheet strengths. However, the determination of $\gamma-iq$ from (\ref{eqn:inv}) is the solution to a class of inverse problems, which are usually ill-conditioned. For this reason we do not pursue this option further.

\begin{figure}
\begin{center}
\begin{minipage}{0.7\linewidth}\begin{center}
\includegraphics[width=0.99\textwidth, angle=0]{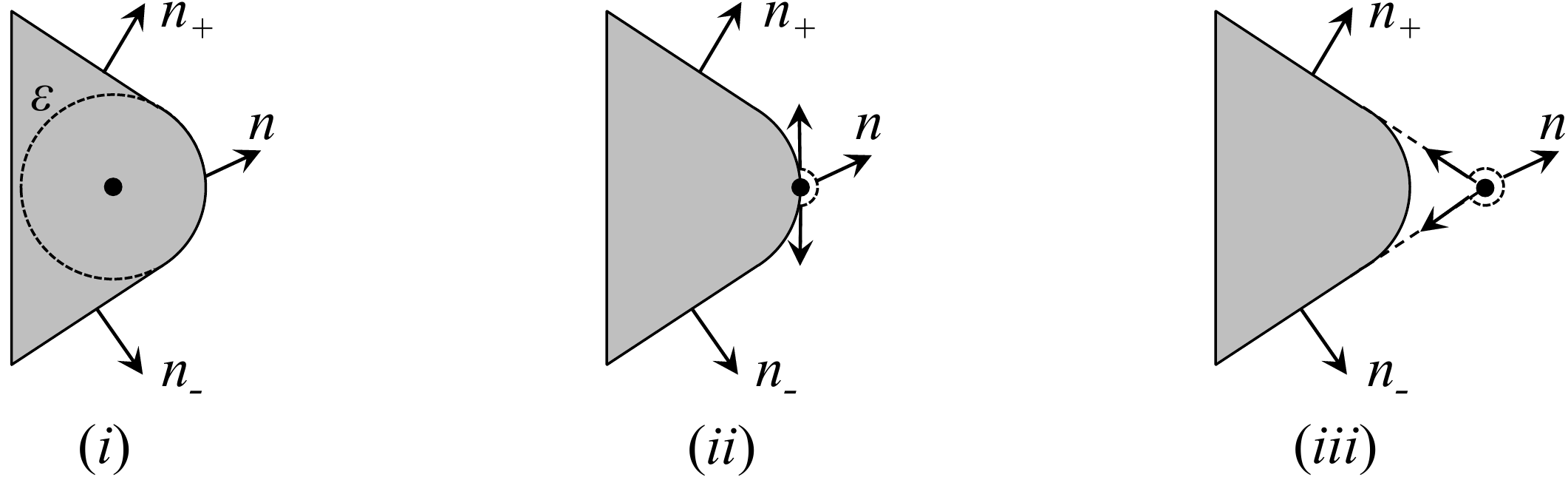}
\end{center}\end{minipage}
\caption{The variation of the surface-normal vector for a rounded corner of radius $\varepsilon\ll 1$ where the black dot represents the location of an irregular point. (\textit{i}) Irregularity inside the surface, (\textit{ii}) irregularity on a smooth boundary, and (\textit{iii}) irregularity on a sharp boundary as $\varepsilon\rightarrow 0$.}
\label{fig:corner}
\end{center}
\end{figure}
\section{Separation at the sharp edge}\label{sec:zetao}
We now shift our focus to applying the vortex-entrainment sheet to the problem of separation at a sharp edge $\vect{x}_o=\zeta_o$. While the normal vector $\uv{n}$ is uniquely defined on the wedge faces, $\vect{x}\neq\vect{x}_o$, it is multivalued at the sharp edge. In order to have a well-posed problem this ambiguity must be removed. To see this, consider a wedge of interior angle $\beta\pi$ with a rounded off corner of radius $\varepsilon\ll 1$ and with an irregular point existing somewhere in the domain as in figure~\ref{fig:corner}. When the irregularity remains inside the surface (case \textit{i}), $\uv{n}$ varies continuously between the two limits $\uv{n}_{\pm}$. When the irregularity is on the boundary of the `dulled' edge (case \textit{ii}), this point must be omitted to maintain an analytic fluid domain. As such, $\uv{n}$ varies through values with a total change in argument of $\pi$ and may have a direction outside those of $\uv{n}_{\pm}$. As $\varepsilon\rightarrow 0$ the irregularity and geometric singularity merge (case \textit{iii}), and not only does $\uv{n}$ become ambiguous, but the change in argument is now $2\pi-\beta\pi$. Hence, the boundary normal vector at the sharp edge, or equivalently the shedding angle, must be defined such that (\ref{eqn:BCv}) is compatible with the interaction of the flows approaching the corner. This process is governed by the intrinsic flow inside the sheets as is shown next. Unlike the case of a cusped edge the shedding angle for the wedge geometry will not, in general, be tangential to either face.

\begin{figure}
\begin{center}
\begin{minipage}{0.32\linewidth}\begin{center}
\includegraphics[width=0.99\textwidth, angle=0]{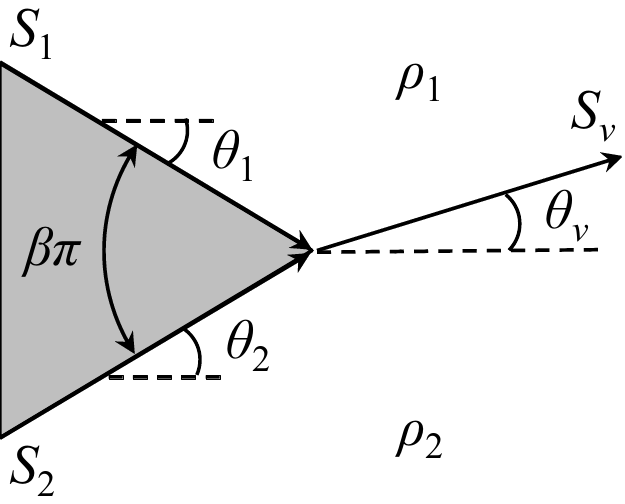}\\(\textit{a})
\end{center}\end{minipage}
\hspace{50pt}
\begin{minipage}{0.36\linewidth}\begin{center}
\includegraphics[width=0.99\textwidth, angle=0]{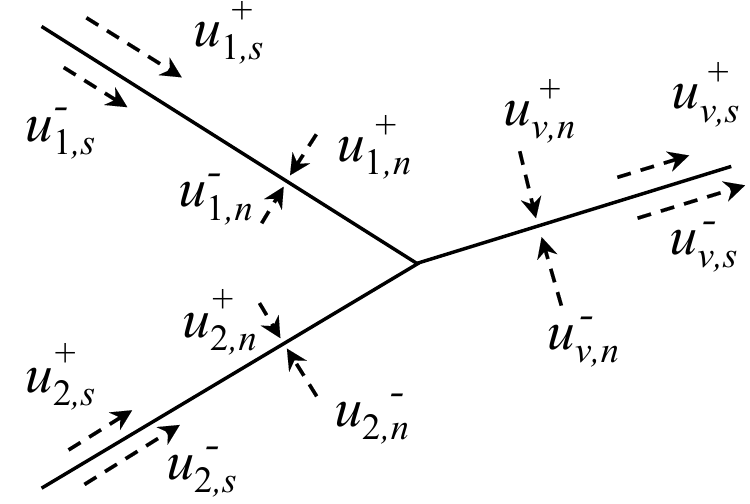}\\(\textit{b})
\end{center}\end{minipage}
\caption{(\textit{a}) Schematic definition of the sheets comprising the wedge of interior angle $\beta\pi$. The directions of the arrows indicate the direction of positive integration in (\ref{eqn:BRE}). (\textit{b}) The left and right limits of the tangential and normal velocities of each sheet. The velocities exist at the sharp edge but are shown displaced from it for clarity.}
\label{fig:wedge}
\end{center}
\end{figure}

The sheet $S_i$ is divided in two: $S_1$ and $S_2$ representing the wedge faces as depicted figure~\ref{fig:wedge}(\textit{a}). The subscripts $(1)$, $(2)$ and $(v)$ will refer to quantities associated with $S_{1}$, $S_{2}$ and $S_{v}$, respectively. The complex sheet strengths are then $\chi_{1}$, $\chi_{2}$ and $\chi_v$, and the arguments of these sheets are $\theta_{1}$, $\theta_{2}$ and $\theta_v$. The entrainment strength is known everywhere except at the sharp edge because the normal direction there is as yet undetermined. Imposing the normal boundary condition on the wedge faces will yield the corresponding vortex sheet strengths $\gamma_1$ and $\gamma_2$. It is assumed that the source of the vorticity in the free sheet $S_v$ results solely from the merging/shedding of the two bound sheets $S_1$ and $S_2$. Hence, the remaining unknowns to be determined are $\gamma_v$, $q_v$ and $\theta_v$.

First, we address the potential outer flow quantities relating to $\vect{u}$ and then discuss the flow $\vect{v}$ inside the sheets. By definition the sheets are characterized by the limiting values just outside the sheets. Hence, we have constraints to impose to ensure the consistency of $S_1$ and $S_2$ merging into $S_v$. Figure~\ref{fig:wedge}(\textit{b}) shows the tangential and normal components of the left and right limit velocities for each sheet. Next, we define $w_v^{\pm}=(u_{v,s}^{\pm}-iu_{v,n}^{\pm})\e{-i\theta_v}$ as the velocities on either side of $S_v$. While $w$ (i.e. $\vect{u}$) may be discontinuous \textit{across} a sheet it is still required to be smooth when $z\rightarrow \zeta=z^{\pm}\in D$ where $\zeta\in\partial D$ since $w$ is sectionally holomorphic \citep{MuskhelishviliNI:46a}. This means that the $w_v^{\pm}$ are related to the velocities just outside of $S_1$ and $S_2$ as:
\begin{eqnarray}
w_{v}^{+}\e{i\theta_v}=(u_{s,1}^{+}-iu_{n,1}^{+})\e{-i(\theta_1-\theta_v)},\quad\quad w_{v}^{-}\e{i\theta_v}=(u_{s,2}^{-}-iu_{n,2}^{-})\e{-i(\theta_2-\theta_v)}.
\end{eqnarray}
The no-slip and no-penetration boundary conditions on the solid wedge faces can now be imposed since $u_{s,1}^{-}$, $u_{s,2}^{+}$, $u_{n,1}^{-}$ and $u_{n,2}^{+}$ will be known from the velocity of the wedge. Adding these components to the above equations, then subtracting and taking real and imaginary parts yields:
\begin{eqnarray}
\gamma_v=u_{s,v}^{-}-u_{s,v}^{+}&=&\gamma_1\cos(\theta_{1}-\theta_{v})+\gamma_2\cos(\theta_{2}-\theta_{v}) \nonumber \\
&&-\left[q_1\sin(\theta_{1}-\theta_{v})+q_2\sin(\theta_{2}-\theta_{v})\right] \label{eqn:gv} \\
q_v=u_{n,v}^{-}-u_{n,v}^{+}&=&\gamma_1\sin(\theta_{1}-\theta_{v})+\gamma_2\sin(\theta_{2}-\theta_{v}) \nonumber \\
&&+\left[q_1\cos(\theta_{2}-\theta_{v})+q_2\cos(\theta_{2}-\theta_{v})\right]. \label{eqn:qv}
\end{eqnarray}
That $\gamma_v$ depends on $q_1/q_2$ and $q_v$ on $\gamma_1/\gamma_2$ is expected from the coupled sheet strength evolution equations in (\ref{eqn:evo}). In particular, we expect that $\theta_1\leq\theta_v\leq\theta_2$ and since $q\geq0$, then entrainment of \textit{irrotational} fluid acts to decrease the vortex sheet strength of the shed sheet by `diffusing' the previously existing vorticity in the sheet. Using the definition of the complex sheet strength in (\ref{eqn:Plemelj}) these equations can be repackaged as:
\begin{equation}\label{eqn:Kutta}
\chi_v=\chi_1+\chi_2.
\end{equation}
In Appendix C it is shown that this requirement is precisely the condition that removes the singularities in the velocity induced at the sharp edge. Hence, (\ref{eqn:Kutta}) might as well be called the Kutta condition, but we note that it derives from the balance of normal momentum and the pressure jump as discussed in \S\ref{sec:evo} and referred to as the `neutralization of the singular pressure gradient' in \S\ref{sec:Kutta}.

Now consider the intrinsic flows $\vect{v}$ inside the sheets. Here we use vector notation to reiterate that $\vect{v}$ is not part of the outer flow, but is confined to the sheet embedded within this flow. When the \textit{tangential} flows $v_{s,1}$ and $v_{s,2}$ in the bound sheets merge at the edge point there must be no normal \textit{impulse} relative to the shedding free sheet $S_v$. For otherwise there would be flow \textit{out} of the sheet. The impulse represents the instantaneous or `ballistic' merging of $S_1$ and $S_2$. This dynamic condition is expressed as:
\begin{equation}\label{eqn:impulse}
0=\left(\left[\rho_{s,1}v_{s,1}\right]\uv{s}_1+\left[\rho_{s,2}v_{s,2}\right]\uv{s}_2\right)\cdot\uv{n}_v.
\end{equation}
Noting from figure~\ref{fig:wedge}(\textit{a}) that $\theta_2=\beta\pi/2$ and $\theta_1=-\beta\pi/2$, then (\ref{eqn:impulse}) yields an expression for the shedding angle as:
\begin{eqnarray}
\tan\theta_v &=& A\tan(\beta\pi/2),\quad\quad\quad A=\left(\frac{\rho_{s,2}v_{s,2}-\rho_{s,1}v_{s,1}}{\rho_{s,2}v_{s,2}+\rho_{s,1}v_{s,1}}\right). \label{eqn:thetav}
\end{eqnarray}
Next, we have the mass and momentum conservation conditions of the merging process, which state that the $\rho_s$ and $\rho_sv_s$ from each of the bound sheets are carried into the free sheet by their respective convective velocities $v_s$:
\begin{eqnarray}\label{eqn:merge}
\rho_{s,v}v_{v,s}=\rho_{s,1}v_{s,1}+\rho_{s,2}v_{s,2}, \quad\quad \rho_{s,v}v_{s,v}^2 = \left(\left[\rho_{s,1}v_{s,s}^2\right]\uv{s}_1+\left[\rho_{s,2}v_{s,2}^2\right]\uv{s}_2\right)\cdot\uv{s}_v.
\end{eqnarray}
With the intrinsic flow quantities $\rho_{s,j}$ and $\vect{v}_j$ ($j=1,2$) known from the solution of the coupled system, then (\ref{eqn:gv}), (\ref{eqn:qv}) and (\ref{eqn:thetav}) represent three equations for the three unknowns $\gamma_v$, $q_v$ and $\theta_v$. Then (\ref{eqn:merge}) gives boundary conditions to calculate the flow into $S_v$ as it is shed.

\subsection{Discussion on special cases of entrainment}\label{sec:edisc}
In this subsection we discuss some sets of simplified circumstances pertaining to separation at a sharp edge that are of practical interest. When $\beta=0$ then (\ref{eqn:thetav}) simply gives $\theta_v=0$ irrespective of the intrinsic flow $\vect{v}$ and thus $\jb{p}$. Of course this corresponds to the usual tangential shedding result for a cusped edge. 

Now consider zero entrainment everywhere so that there is no intrinsic flow $\vect{v}=0$ or mass $\rho_s=0$ in the sheets, namely the conventional vortex sheet. Equation (\ref{eqn:thetav}) then becomes indeterminate and a new condition for the shedding angle is required. This is obtained from (\ref{eqn:qv}), which with each $q_j=0$ reduces to:
\begin{equation}
\gamma_1\sin(\beta\pi/2+\theta_v)=\gamma_2\sin(\beta\pi/2-\theta_v).
\end{equation}
There is the trivial solution with $\gamma_1=\gamma_2=0$ meaning no flow at the edge. The only other solutions that put $\theta_v$ in a physical range are $\theta_v=\pm\beta\pi/2$ and with the respective consequences of $\gamma_1=0$ or $\gamma_2=0$. This corresponds to a tangential shedding angle and that the flow is stagnated on one side of the vortex sheet depending on the sign of shed circulation. Hence, as mentioned in \S\ref{sec:Kutta} the vortex sheet from one wedge face convects off the surface with no contribution from the opposing face. These features are consistent with the results of \cite{Pullin:78a} in which there was zero entrainment.

Lastly, consider the onset of motion at $t=0$ where no vorticity has been shed from the sharp edge and $S_v$ does not yet exist. As used by \cite{RottN:56a} and \cite{Pullin:78a} the attached potential flow responsible for separation is obtained as the leading order term in an expansion of the complex potential near the sharp edge: $\Phi\sim (z-\zeta_o)^n$ where $n=1/(2-\beta)$ for the wedge of angle $\beta\pi$. The imposed pressure gradient at infinity that drives this flow suggests the windward wedge face will `feel' the pressure at $t=0^+$ before the leeward face. Hence, we would expect more significant entrainment to occur on the windward face, say $S_2$, so that $v_{2,s}\gg v_{1,s}$ and (\ref{eqn:thetav}) gives $A\approx 1$ and $\theta_v\approx\beta\pi/2$. This is again consistent with the small-time similarity solution. In a real flow the ensuing roll-up would be sure to quickly induce a pressure and so entrainment on the leeward face thus changing $\theta_v$.

\section{Example calculations}\label{sec:examps}
This section presents two example calculations to demonstrate the vortex-entrainment sheet concept as applied to separation at a sharp edge. For simplicity we make some further assumptions on the dynamics for a \textit{free} sheet. This is accomplished by enforcing a relation between $\vect{v}$ and $\vect{u}$. To this end, we recognize that the conventional vortex sheet has the Birkhoff-Rott equation as its evolution equation, which says the sheet moves with the average fluid velocity. Motivated by this, we simply set $\vect{v}=\overline{\vect{u}}$ and we then have the generalized Birkoff-Rott equation (\ref{eqn:BRE}) to convect the free sheet.

To see the dynamical consequences of our assumption we apply them to the mass and momentum equations (\ref{eqn:mass_3D})-(\ref{eqn:momentum_3D}) to give:
\begin{eqnarray}\label{eqn:mass_mom_simp}
\pder{\rho_s}{t}+\pder{}{s}\left(\rho_sv_s\right)=\rho q,\quad\quad 
\rho_s\Der{_sv_s}{t}=\jb{\tau}, \quad\quad
\rho_s\Der{_sv_n}{t}=-\jb{p} 
\end{eqnarray}
We see that the normal momentum equation remains unchanged regardless of the value of the entrainment strength $q$. Hence, the pressure jump is exactly that required to move the mass in the sheet with $v_n=\overline{u_n}$. For a free sheet we have $\tau^{+}=\tau^{-}=0$ since the outer flow is potential and thus $\jb{\tau}=0$ means the tangential motion of particles in the sheet move materially. However, $\jb{\tau}$ could be retained as a way to model a mixing process of the intrinsic sheet flow $v_s$.

\subsection{Setup of computed simulations}
First we present the problem setup for a case with \textit{zero} entrainment that is given in \S\ref{sec:Pullin}. A case with non-zero entrainment is then considered in \S\ref{sec:osc} and the required modifications to the problem setup are also discussed there. The geometry of each problem is a wedge of infinite extent. 

We use a combination of the Laplace and boundary integral formulations. The wedge is stationary and a boundary condition is imposed at infinity to drive the flow in the form of a known harmonic potential function $\phi_\infty(r,\theta,t)=\text{Re}\left\{\Phi_\infty(z,t)\right\}$ that represents the attached flow. Since there is no entrainment $S_1$ and $S_2$ have no-penetration conditions and $S_v$ has a continuity of normal velocity condition. Moreover, by construction the potential $\phi_\infty$ satisfies homogeneous Neumann boundary conditions on the wedge faces. Hence, after each new free vortex sheet segment is shed, we need only to account for the normal velocity induced by $S_v$ on the walls. We can then arrange the problem for a potential $\phi_w$, which represents the image system of the free vortex sheet inside the wedge. This $\phi_w$ satisfies:
\begin{eqnarray}
\nabla^2\phi_w=0 \quad\quad &\text{for}& \quad r> 0,\quad \theta\in[-\theta_w,\theta_w] \label{eqn:phiw}\\ 
\frac{1}{r}\pder{\phi_w}{\theta}=-\text{Re}\left\{i\e{i\theta}w_v(r,\theta)\right\}\quad\quad &\text{for}& \quad r> 0,\quad \theta=\pm\theta_w \label{eqn:wall} \\
\phi_w \rightarrow 0\quad\quad &\text{for}& \quad r\rightarrow \infty,\quad \theta\in[-\theta_w,\theta_w] \label{eqn:inf}
\end{eqnarray}
where $\theta_w=\pi(1-\beta/2)$ and ($\pm$) corresponds to the angular coordinates of $S_1$ and $S_2$, and so $w_v(r,\pm\theta_w)$ is the velocity induced by $S_v$ on the wedge faces. For brevity we will write $\zeta_w=r\e{\pm i\theta_w}$ for the coordinate on either $S_1$ or $S_2$.

In \S\ref{sec:edisc} we saw that for zero entrainment the shedding angle equation (\ref{eqn:thetav}) becomes indeterminate and $\theta_v$ is instead obtained by (\ref{eqn:qv}) with $q_v=q_1=q_2=0$. Specifically $\theta_v$ is necessarily tangent to one wedge face and the flow is stagnated at the apex on the other face. This second condition could be imposed by $\partial\phi/\partial r=0$ on the $S_1$ side where $\phi$ is the total potential. However, it is automatically satisfied by imposing $\theta_v\equiv \beta\pi/2$. We note that this is a result of having to impose the no through-flow condition on both $S_1$ and $S_2$ as well as the continuity of normal velocity on each side of $S_v$, all of which exist at the same location. As a consequence of all these constraints, the strength of the new vortex sheet segment of $S_v$ is $\gamma_v=\gamma_2$ by (\ref{eqn:gv}), which represents $S_2$ convecting off to form $S_v$. 

The calculation procedure is as follows. Given at time $t_0=0$ is the vortex sheet strength $\gamma_w$ on $S_1$ and $S_2$ from $\phi_\infty(\zeta_w,t_0)$ and an initialized position $\zeta_v$ and vortex sheet strength $\gamma_v$ of the free sheet $S_v$. The flow is initialized by representing the first shed segment of $S_v$ as a point vortex as was similarly done by \cite{JonesMA:03a}. Then, for each $k\geq 1$ time step:
\begin{enumerate}
\item Compute new velocity on wedge faces due to $S_v$ as
\begin{equation}
w_v(\zeta_w)=\frac{1}{2\pi i}\int_{S_v}\frac{\gamma_v(s)\md{s}}{\zeta_w-\zeta_v(s)}.
\end{equation}
\item Solve the Laplace equation for $\phi_w$ in (\ref{eqn:phiw}) with boundary conditions (\ref{eqn:wall})-(\ref{eqn:inf}).
\item Compute new $\gamma_w$ along $S_1$ and $S_2$ (i.e. $\gamma_1(r)$ and $\gamma_2(r)$) as:
\begin{equation} 
\gamma_w(\zeta_w)=\pder{}{r}(\phi_w(\zeta_w)+\phi_\infty(\zeta_w))+\text{Re}\left\{w_v(\zeta_w)\e{i\theta(\zeta_w)}\right\}. 
\end{equation}
\item Set $\gamma_v=\gamma_2(r=0)$ and $\theta_v=\beta\pi/2$ of new sheet segment of $S_v$. 
\item Compute new total induced velocity on free sheet $S_v$ as
\begin{equation}
w(\zeta_v)=\pder{\overline{\zeta_v}}{t}=\der{\Phi_\infty}{z}+\frac{1}{2\pi i}\int_{S_v}\frac{\gamma_v(s)\md{s}}{\zeta_v-\xi_v(s)}+\frac{1}{2\pi i}\int_{S_1\cup S_2}\frac{\gamma_w(s)\md{s}}{\zeta_v-\zeta_w(s)}.
\end{equation}
\item Time integrate $\partial\overline{\zeta_v}/\partial t$ to advect free sheet $S_v$ to $t_{k+1}$.
\end{enumerate}
\vspace{2pt}
In steps (v) and (vi) $\overline{\zeta}_v$ is the complex conjugate of the sheet position. Due to the simple geometry we use the method of images that maps the domain outside the wedge to a semi-infinite plane as $z^*=z^n$. This combines steps (i)-(iii) to obtain $\gamma_w$ directly. The integrals may be computed with a discrete sheet method or as a system of point vortices. Lastly, the free sheet $S_v$ is advected with a fourth-order Runge Kutta scheme.

\begin{figure}
\begin{center}
\begin{minipage}{0.27\linewidth}\begin{center}
\includegraphics[width=0.99\textwidth, angle=0]{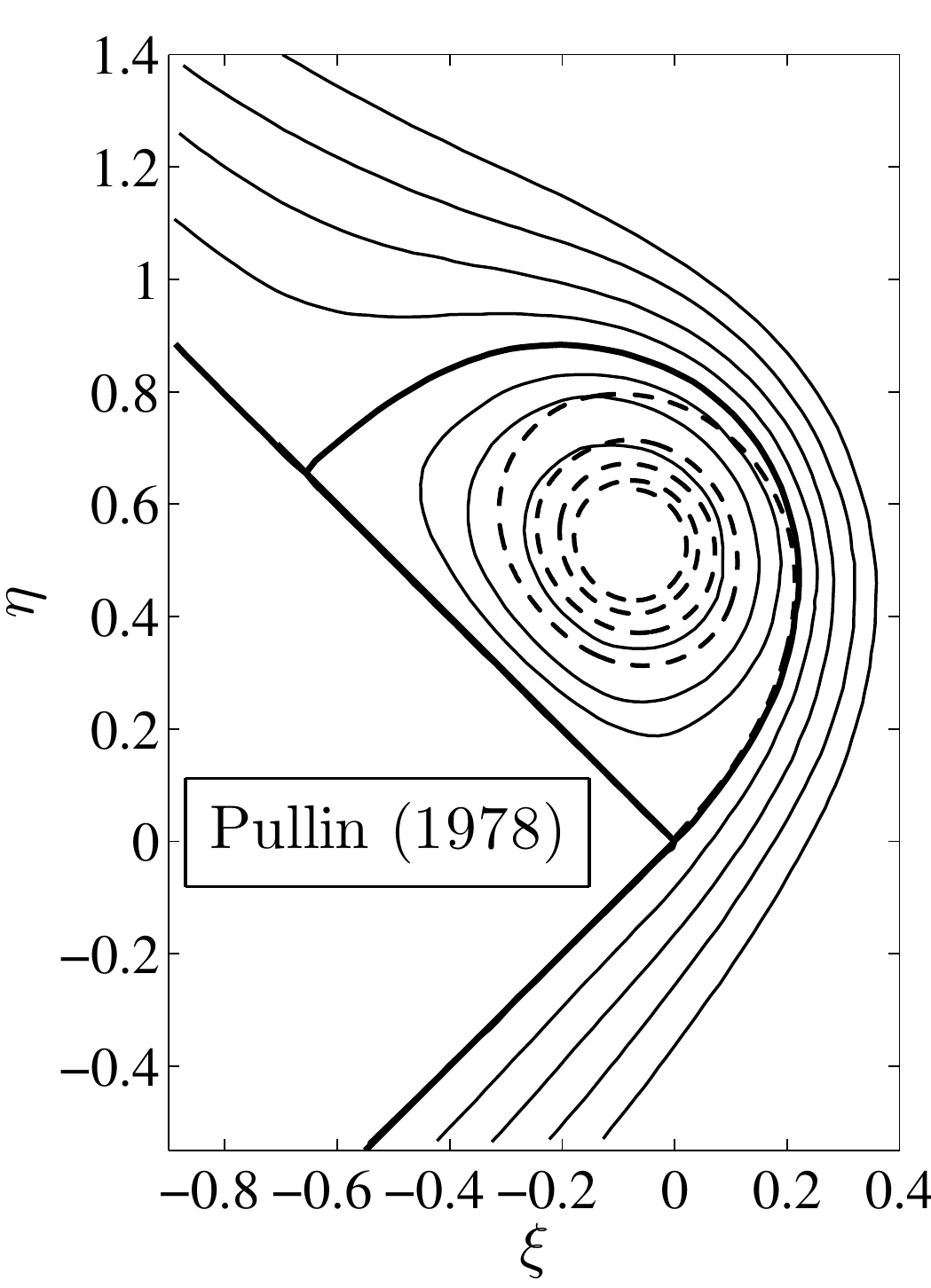}
\end{center}\end{minipage}
\begin{minipage}{0.27\linewidth}\begin{center}
\includegraphics[width=0.99\textwidth, angle=0]{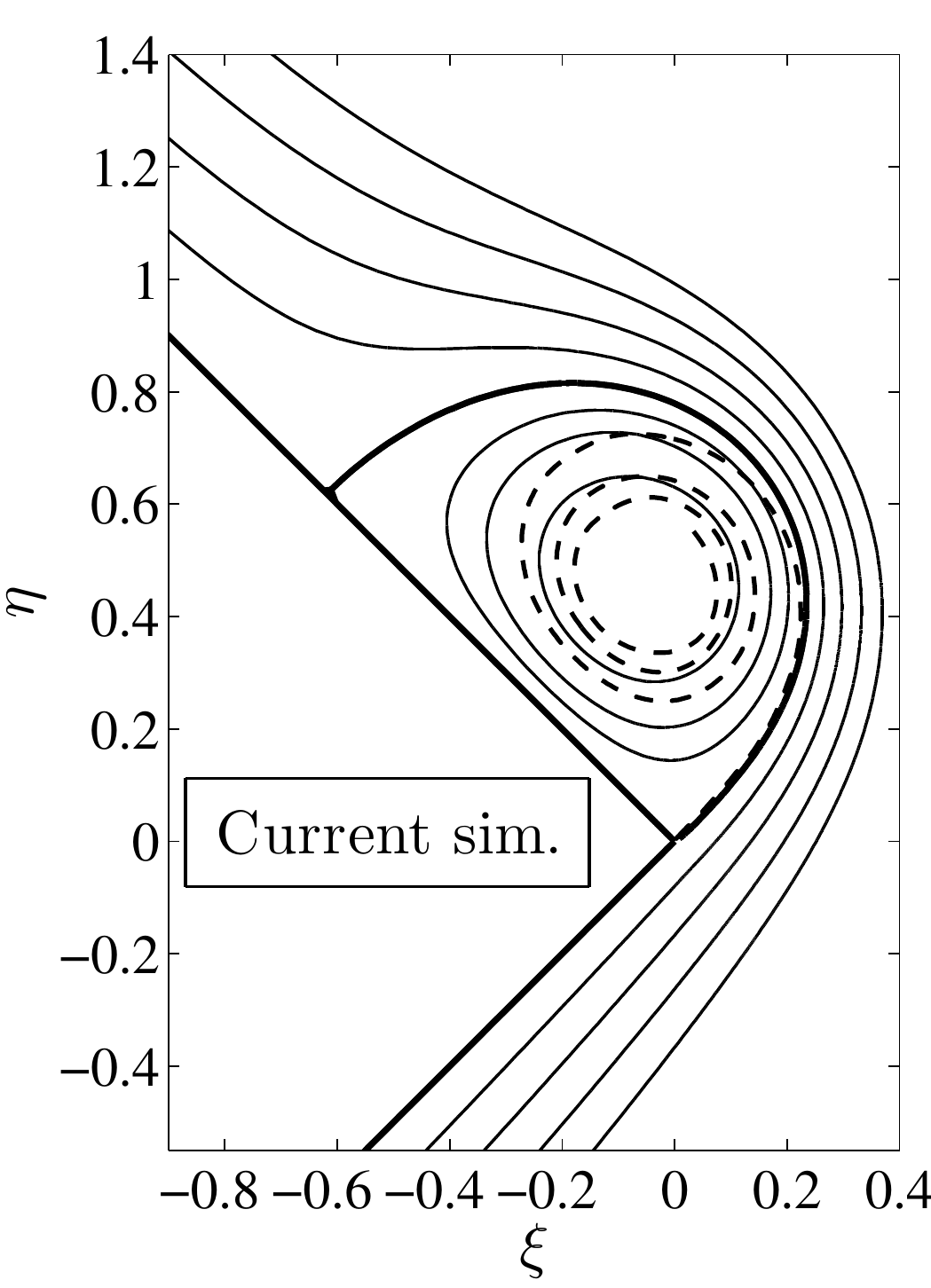}
\end{center}\end{minipage}
\begin{minipage}{0.4\linewidth}\begin{center}
\includegraphics[width=0.99\textwidth, angle=0]{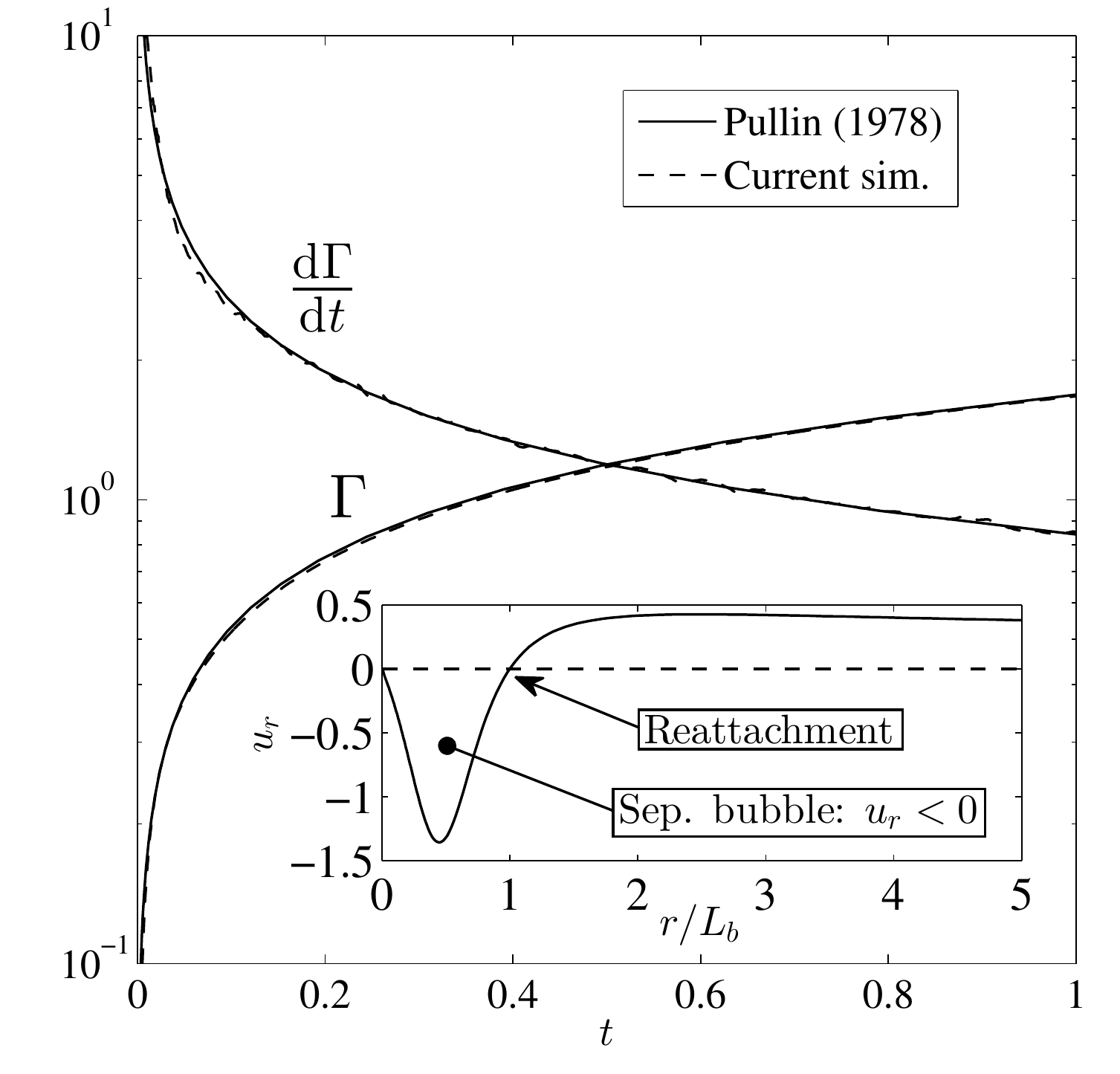}
\end{center}\end{minipage}
\begin{minipage}{0.54\linewidth}\begin{center}
(\textit{a})
\end{center}\end{minipage}
\begin{minipage}{0.4\linewidth}\begin{center}
(\textit{b})
\end{center}\end{minipage}
\caption{(\textit{a}) Comparison of streamlines (solid) and vortex sheet position (dashed) for the case with $m=0$ and $\beta\pi=\pi/2$ ($=90^\circ$) at $t=1$ scaled to the simlarity space $\omega=\xi+i\eta$. (\textit{b}) The total and rate of shed circulation as a function of time. The inset shows the radial velocity along the upper wall, $u_r$, as a function of distance from the apex normalized by the separation bubble length $L_b$. The flow is stagnated at the apex and the reattachment point $r=L_b$ with reversed flow in the separation bubble.}
\label{fig:Pullin}
\end{center}
\end{figure}
\subsection{Starting flow past an infinite wedge}\label{sec:Pullin}
As validation we first reproduce a result from the similarity solutions of \cite{Pullin:78a} for a starting flow over an infinite wedge of interior angle $\beta\pi$. There is zero entrainment everywhere for all time and the attached flow complex potential is $\Phi_\infty(z,t)=-iat^mz^n$ where $n=1/(2-\beta)$. We simulate a time-dependent flow beginning from $t=0$. 

Figure~\ref{fig:Pullin}(\textit{a}) shows a comparison of the streamlines and vortex sheet location for the case with $m=0$ and $\beta\pi=\pi/2$. The computed simulation is at $t=1$ and is scaled to the similarity space $\omega=\xi+i\eta$ \citep[for more detail see][]{Pullin:78a}. There is good agreement between the visual character of the flow field. For a more quantitative comparison figure~\ref{fig:Pullin}(\textit{b}) plots the total and rate of shed circulation, again showing excellent agreement with the similarity laws given by Pullin. The inset shows the velocity along the upper wedge surface (i.e. $S_1$), which stagnates at the leeward side of the apex as expected from the boundary condition for the case of zero entrainment.

\subsection{Oscillating flow with estimated entrainment}\label{sec:osc}
Technically the entrainment strengths are to be given as boundary conditions. However, we may estimate the entrainment $q_v$ of the free sheet by relaxing the condition of stagnated flow on the leeward side of the apex. In other words, a non-zero wall-tangential velocity is allowed just outside both $S_1$ and $S_2$ at the apex: $\gamma_1\neq0$ and $\gamma_2\neq0$. Each of these velocities then has a normal component relative to the free sheet and must merge to form $S_v$. Hence, the merging of the $S_1$ and $S_2$ sheets is a form of entrainment. This means a non-tangential shedding angle is possible and this is determined next.

These wall-tangent flows are in the outer potential flow $\vect{u}$ just upstream of the sharp edge point and thus must be entrained through the sharp edge to become the intrinsic flow $\vect{v}$. Therefore, since there cannot be flow through $S_v$ the normal impulse relative to $S_v$ of these incoming flows must be equal just as was the case for (\ref{eqn:impulse}). Hence, we obtain an analogous equation for the shedding angle:
\begin{equation}\label{eqn:thetavB}
\tan\theta_v=B\tan(\beta\pi/2)\quad\quad B=\left(\frac{u_{s,2}^{-}-u_{s,1}^{+}}{u_{s,2}^{-}+u_{s,1}^{+}}\right).
\end{equation}
Although we have assumed $\jb{\rho}=0$ we could accommodate a stratified flow by the substitution $u^{\pm}\rightarrow\rho^{\pm}u^{\pm}$. The above equation is equivalent to the one used by \cite{Mohseni:17m}; the two relations are related by the trigonometric identity $\arctan(x)=\arccos(1/\sqrt{1+x^2})$. However, there it was not recognized that this necessarily requires entrainment into the shed sheet. 

\begin{figure}
\begin{center}
\begin{minipage}{0.5\linewidth}\begin{center}
\hspace{45pt}
\begin{minipage}{0.7\linewidth}\begin{center}
\includegraphics[width=0.99\textwidth, angle=0]{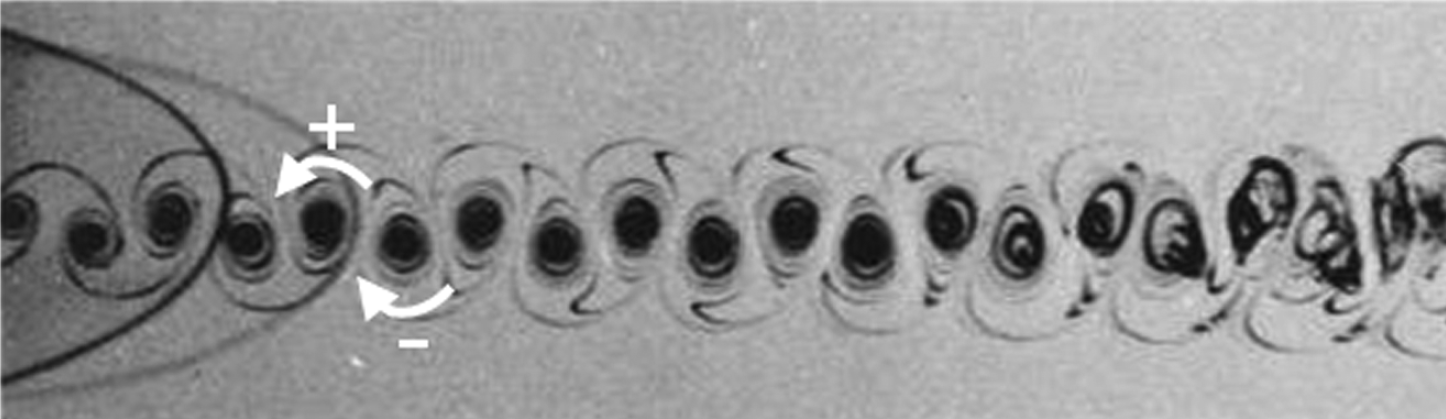}
\end{center}\end{minipage}
\\
\begin{minipage}{0.99\linewidth}\begin{center}
\includegraphics[width=0.99\textwidth, angle=0]{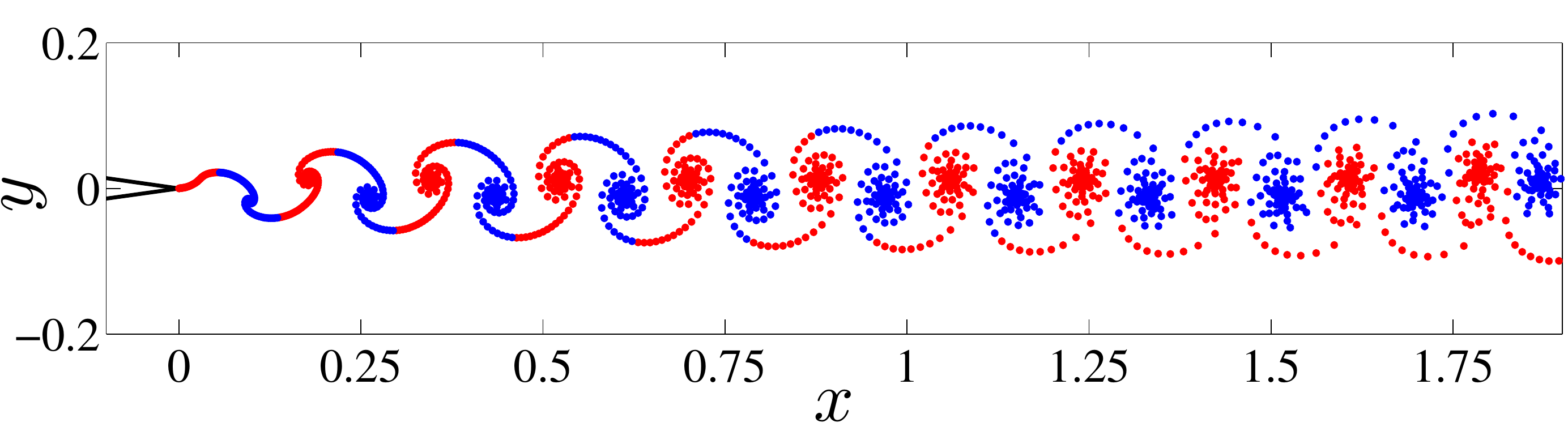}
\end{center}\end{minipage}
\end{center}\end{minipage}
\begin{minipage}{0.49\linewidth}\begin{center}
\includegraphics[width=0.99\textwidth, angle=0]{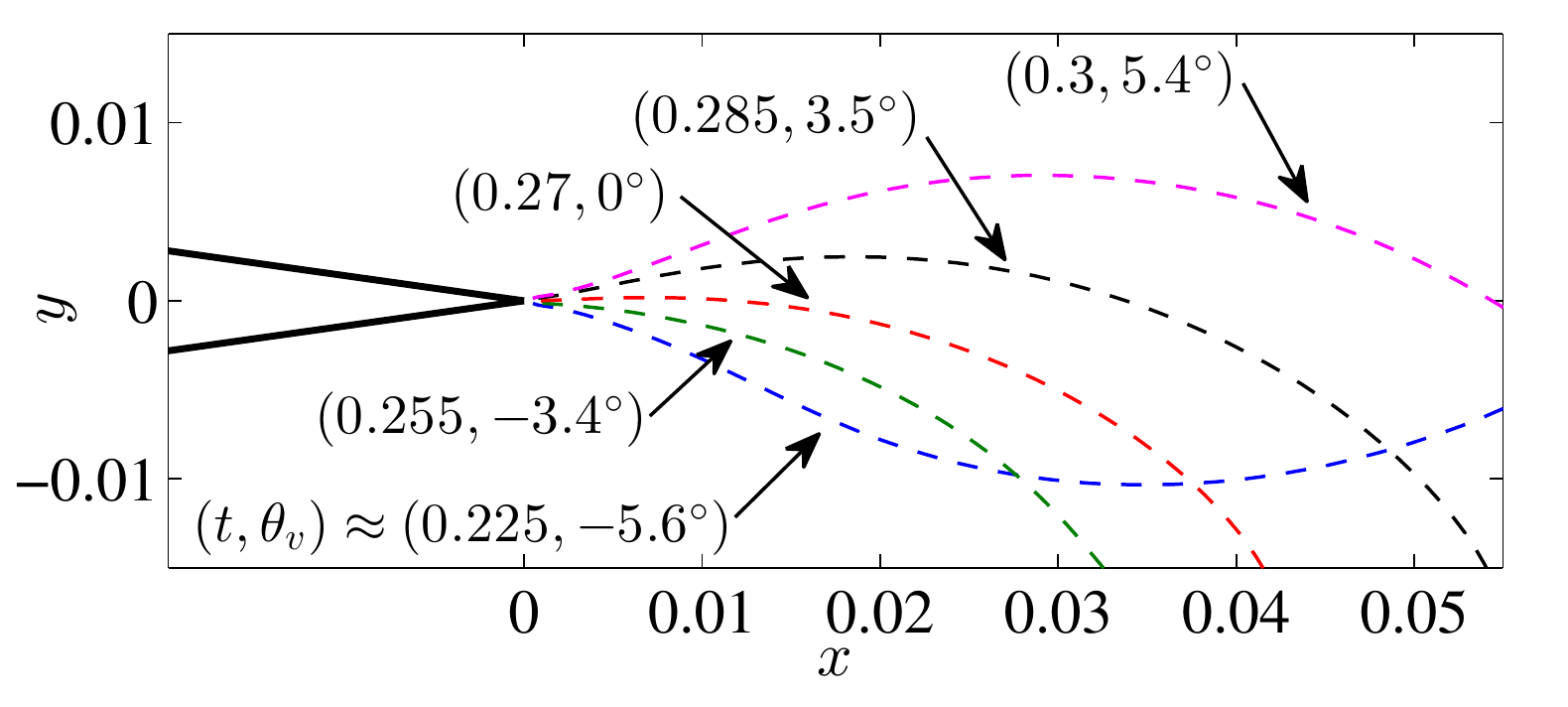}
\end{center}\end{minipage}
\begin{minipage}{0.5\linewidth}\begin{center}
\begin{center}
(\textit{a})
\end{center}
\end{center}\end{minipage}
\begin{minipage}{0.49\linewidth}\begin{center}
\begin{center}
(\textit{b})
\end{center}
\end{center}\end{minipage}
\caption{(\textit{a}) Wake of an oscillatory trailing edge flow. (\textit{Top}) Experimental flow visualization of a NACA 0012 airfoil oscillating at $f=6$ Hz from \cite{Koochesfahani:89a}. (\textit{Bottom}) Qualitative comparison of a simulated wake at $t=2$ with wedge angle $16^\circ$ to approximate the airfoil trailing edge. (\textit{b}) Zoomed in view of the sharp edge showing the sheet shapes at different times $t$ with corresponding shedding angles $\theta_v(t)$ as labeled with $(t,\theta_v)$.}
\label{fig:osc2}
\end{center}
\end{figure}
Since $\theta_v$ is now determined by (\ref{eqn:thetavB}) instead of prescribed as $\beta\pi/2$, then step (iv) of the procedure given in \S\ref{sec:Pullin} must be updated as:
\vspace{3pt}
\begin{itemize}
\item[(iv)$^\prime$] Compute $\gamma_v$ and $q_v$ from (\ref{eqn:gv}) and (\ref{eqn:qv}) from $\gamma_1$, $\gamma_2$, $\theta_v$ with $q_1=q_2=0$.
\end{itemize}
\vspace{3pt}
In the same way that a segment of vortex sheet must be of finite length to possess circulation, $\Gamma=\int\gamma\md{s}$, so too must an entrainment sheet segment have a finite length to have an entrainment rate, $Q=\int q\md{s}$. By confining the entrainment over some small sheet segment $\Delta s$ to the point at the sharp edge means we are essentially approximating a net entrainment rate $Q$ as
\begin{equation}
Q=\int_{\Delta s} q_v\md{s}\approx q_v\Delta s.
\end{equation}
Then the amount of mass of fluid with density $\rho$ that is entrained is $\Delta m\approx \rho Q\Delta t$. Hence, the sheet mass density $\rho_{v,s}=\Delta m/\Delta s$ put into $S_v$ in a given amount of time is:
\begin{equation}
\rho_{v,s}\approx \rho q_v \Delta t, 
\end{equation}
which is equivalent to integrating the mass equation in (\ref{eqn:mass_mom_simp}) over $\Delta t$. Therefore, with a constant $\Delta t$ and $\rho$ the behavior of $\rho_{v,s}$ and $q_v$ are synonymous with the understanding that $q_v$ exists only at the edge and puts $\rho_{v,s}$ in the sheet which then convects downstream. In other words, the time trace of $q_v$ is analogous to the trace of $\rho_{v,s}$ along the sheet arclength at a given time.

Having physically explained the estimated entrainment through the sharp edge, we now compute a solution that is similar to the wake created by the trailing edge of an oscillating airfoil. The driving potential flow is given by 
\begin{equation}
\Phi(z,t)=A_1(t)(z-\zeta_o)^n+A_2(t)(z-\zeta_o)^{2n},
\end{equation}
where $A_1(t)=A_o\cos(2\pi ft)$ is the oscillating component and $A_2=U_\infty$ represents the translational velocity of the airfoil. To provide a qualitative comparison we matched the oscillation frequency from one of the dye visualization experiments of \cite{Koochesfahani:89a}. The experiments used a NACA 0012 airfoil with chord-based Reynolds number $Re_c=11,400$ and $2^\circ$ oscillation amplitude and frequencies $f=4$, 5 and 6 Hz. We ran a simulation with $f=6$ Hz and where $q_v$ was estimated as described above. Due to the more complicated sheet shape, we used a point vortex method to approximate the sheet.

\begin{figure}
\begin{center}
\begin{minipage}{0.7\linewidth}\begin{center}
\includegraphics[width=0.99\textwidth, angle=0]{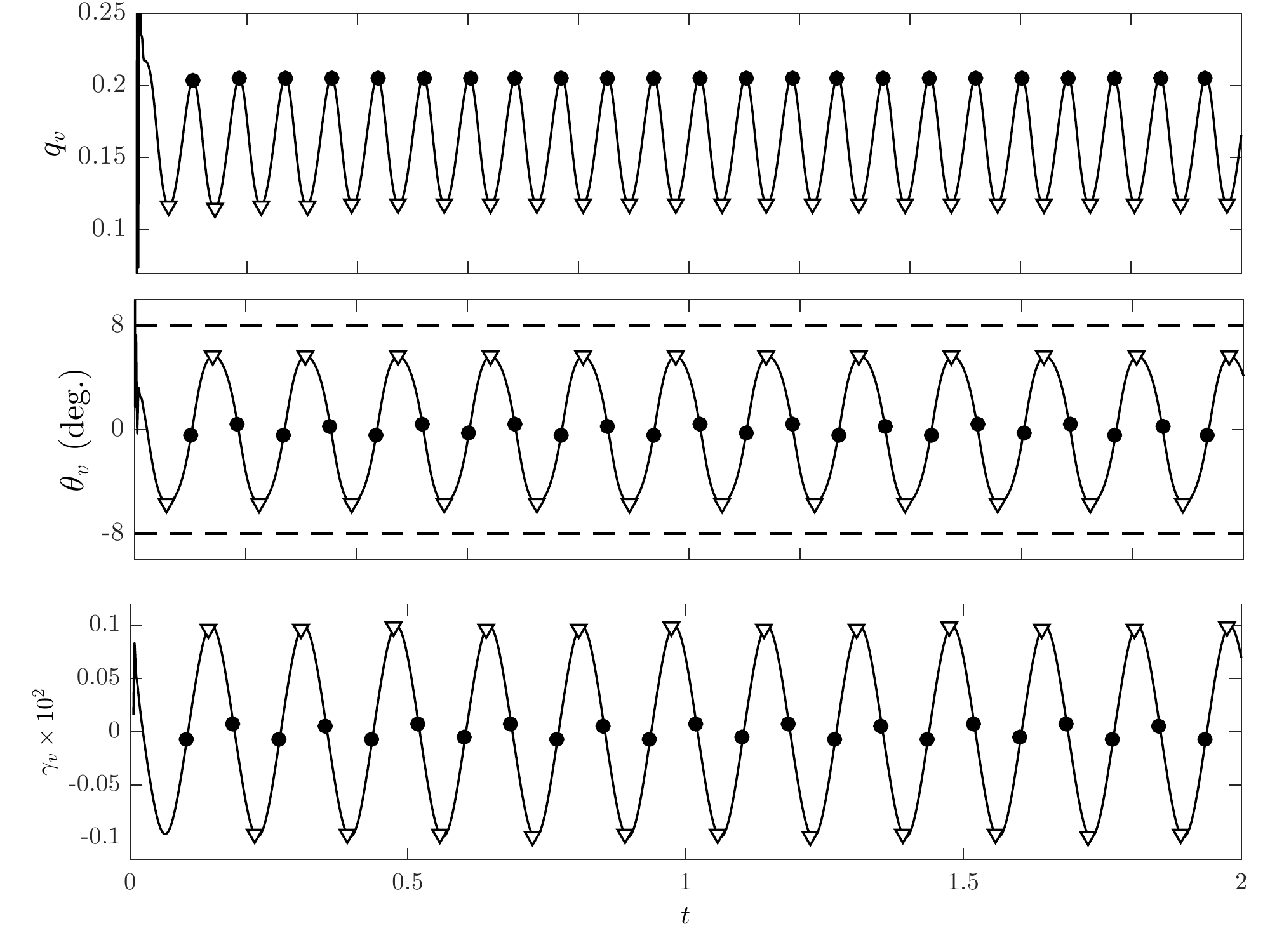}
\end{center}\end{minipage}
\caption{Computed quantities for the simulated flow in figure~\ref{fig:osc2}. (\textit{Top}) The entrainment sheet strength at the sharp edge, (\textit{middle}) shedding angle $\theta_v$ where the dashed lines are the tangential limits $\beta\pi/2=\pm 8^\circ$, and (\textit{bottom}) vortex sheet strength at the edge. All are plotted against time. The solid circles and open triangles respectively mark where the entrainment strength is strongest and weakest.}
\label{fig:osc}
\end{center}
\end{figure}
Figure~\ref{fig:osc2}(\textit{a}) plots a comparison of the simulated wake structure at $t=2$ and shows reasonably good agreement with the experiment. Figure~\ref{fig:osc2}(\textit{b}) shows a zoomed in view of the sharp edge and plots the free sheet shape at several times through a period of oscillation. We can clearly see the smooth variation of the shedding angle $\theta_v$ . The computed entrainment sheet strength $q_v$, shedding angle $\theta_v$ and vortex sheet strength $\gamma_v$ are the top, middle and bottom plots in figure~\ref{fig:osc}, respectively. First, we see that $q_v$ oscillates with twice the frequency as the oscillating flow, but is shifted upward by a positive mean, $\overline{q_v}\approx 0.159$, such that $q_v(t)>0$. This is to be expected since entrainment is a viscous process that cannot be reversed such that the fluid is returned to an irrotational state. The shedding angle $\theta_v$ shows the expected oscillatory behavior within the two tangential shedding limits of $\pm\beta\pi/2$ and about a zero mean, namely the wedge bisector and likewise for the vortex sheet strength.

The solid circles mark when $q_v$ is strongest and the open triangles mark when $q_v$ is weakest. Although we have oscillated the flow, we will now speak as if the trailing edge were oscillating instead. Hence, entrainment is strongest when the trailing edge is passing through the zero incidence position. At these times the edge is moving with its highest velocity and changes sign of \textit{acceleration}. This deceleration is known to cause an impingement of the flow onto the leeward side of the wedge and a corresponding significant normal force \citep{DeVoriaAC:13a}. This large pressure/normal force on the wedge is then responsible for the strong entrainment. The vortex sheet displays the opposite behavior showing maxima and minima when the wedge changes sign of \textit{velocity} at the largest amplitudes of the oscillation. This conjugacy between the vortex sheet and entrainment sheet, or in general tangential and normal quantities, is a feature that we have repeatedly observed throughout this work.

\section{Concluding remarks}
This study proposed the vortex-entraiment sheet as a model of viscous boundary and shear layers in three-dimensional flow. The sheet differs from the conventional vortex sheet by allowing mass and consequently momentum in the sheet. The process of entrainment is the mechanism that allows fluid to enter the sheet, thus endowing it with inertia. Hence, there is an intrinsic flow confined to the tangential manifold defining the sheet position. This internal flow is dynamically coupled to the flow outside the sheet and may also have physical properties and phenomena that are different from the bulk fluid and that exist only on the sheet.

The physical concept motivating the entrainment sheet definition is the preservation of the mass within a finite-thickness viscous layer in the limit as the layer collapses to zero thickness. In other words, the sheet has a mass per-unit-area of sheet, $\rho_s$. The essential characteristics of the sheet are the vectorial vortex sheet strength representing tangential discontinuities in the velocity, while the scalar entrainment sheet strength corresponds to a discontinuity of the normal velocity. The latter sheet is obviously responsible for the entrainment, which acts as a source of $\rho_s$ relative to the flow inside the sheet. 

The velocity field induced by the vortex-entrainment sheet is given by a generalized Birkhoff-Rott equation where the sheet has a complex strength with the real part corresponding to the vortex sheet as usual and the imaginary part to the entrainment sheet. This is similar to the conventional vortex sheet, but the dynamical consequences are different. Namely, the generalized equation is not necessarily the evolution equation since the sheet contains mass. By allowing the sheet to have mass it is able to support a pressure jump. This is a stark difference from the conventional vortex sheet, which remains dynamically indistinct from the surrounding fluid. A major physical consequence of this is that no explicit Kutta condition is required as the flow remains finite due to a balance of normal momentum with this pressure jump. More specifically, for the case of shedding from a sharp edge the shedding angle is dictated by the normal impulse of the intrinsic flows in the sheets that merge at the edge to form the free sheet.

Throughout this work a theme of conjugacy between the vortex sheet/tangential quantities and the entrainment sheet/normal quantities was observed. We interpret this to mean the entrainment sheet is a natural, orthogonal extension of the conventional vortex sheet as a model of viscous layers. The vortex-entrainment sheet concept was applied to a few example calculations and showed encouraging results about representing the viscous phenomenon of entrainment with an inviscid model.

This paper is the first part in a planned two-part series. The second part will extend the vortex-entrainment sheet model to the case of separation on a \textit{smooth} surface where the separation point is able to move relative to the surface. In addition, the problem of predicting the location of the separation point is considered.


\section*{Appendix A: Singular parts of the curl and divergence}
This appendix represents the discontinuity in $\vect{u}$ across the sheet as a Heaviside function to obtain the singular parts of the curl and divergence operators as given in (\ref{eqn:sing_dist}). The following derivation essentially parallels that given by \cite{HoweMS:07a} for a vortex sheet. Near the sheet the discontinuous fluid velocity can be expressed as:
\begin{equation}
\vect{u}=H(n^{+})\vect{u}^{+}+H(n^{-})\vect{u}^{-}
\end{equation}
where $n=0$ specifies the sheet location and $H(n)$ is the Heaviside function with defining property $H(n>0)=-H(n<0)=H(n)$. Substituting this into the curl and divergence of the velocity gives:
\begin{eqnarray}
\vort=\curl{u}=\grad{H(n)}\times\left(\vect{u}^{+}-\vect{u}^{-}\right)\quad\quad\quad
\Delta=\mdiv{u}=\grad{H(n)}\cdot\left(\vect{u}^{+}-\vect{u}^{-}\right)
\end{eqnarray}
since $\vect{u}^{+}$ and $\vect{u}^{-}$ are incompressible, irrotational flows existing in the open region defined by the fluid domain $D$ whose closure contains the sheet. Substituting $\grad{H(n)}=\delta(n)\uv{n}$ gives:
\begin{eqnarray}
\vort=\uv{n}\times\jb{\vect{u}}\delta(n)=\boldsymbol{\gamma}\delta(n)\quad\quad\quad
\Delta=\uv{n}\cdot\jb{\vect{u}}\delta(n)=-q\delta(n).
\end{eqnarray}
where $\boldsymbol{\gamma}(s,b)$ is a sheet-tangent vector giving the strength of the vortex sheet, and $q(s,b)$ is the strength of the entrainment sheet as given by (\ref{eqn:g_3D}) and (\ref{eqn:q}).

\section*{Appendix B: Conjugate definitions}\label{sec:defs}
Let $s$ and $n$ be the tangential and normal sheet coordinates on a one-dimensional sheet immersed in a two-dimensional flow. The amount of circulation within and the flux into a sheet segment are:
\begin{eqnarray}\label{eqn:GQ}
\Gamma=\int\vect{u}\cdot\uv{s}\md{s},\quad\quad\quad
Q=-\int\vect{u}\cdot\uv{n}\md{s}.
\end{eqnarray}
Using $\vect{u}=\grad{\phi}$ and the Cauchy-Riemann relations, the tangential and normal components of velocity, $u_s$ and $u_n$, are 
\begin{eqnarray}
u_s=\pder{\phi}{s}=\pder{\psi}{n},\quad\quad u_n=\pder{\phi}{n}=-\pder{\psi}{s}.
\end{eqnarray}
As such, the integrands in (\ref{eqn:GQ}) can be written as exact differentials of $\phi$ and $\psi$, respectively. Therefore, integrating on each side of the sheet we have:
\begin{eqnarray}
\Gamma=-(\phi^{+}-\phi^{-})=-\jb{\phi},\quad\quad Q=(\psi^{+}-\psi^{-})=\jb{\psi}.
\end{eqnarray}
Taking the partial derivative of these quantities along the tangential coordinate gives the vortex and entrainment sheet strengths, $\gamma$ and $q$, as:
\begin{eqnarray}\label{eqn:gq1}
\gamma=\pder{\Gamma}{s}=u_s^{-}-u_s^{+}, \quad\quad
q=\pder{Q}{s}=u_n^{-}-u_n^{+}.
\end{eqnarray}
It is worth reiterating that $Q>0$, $q>0$ corresponds to entrainment \textit{into} the sheet as signified by the negative sign in the equation defining $Q$. 

\section*{Appendix C: Removal of singularities}
\textit{Comment}: In the theory of singular integral equations the sheet strength is conventionally defined as the jump in $w$, that is $\chi=\jb{w}$ and the Cauchy kernel is $(\zeta-z)^{-1}$. However, in (\ref{eqn:BRE}) and (\ref{eqn:Plemelj}) we defined the sheet strength to be minus the jump so that the kernel is $(z-\zeta)^{-1}$. This was done so as to avoid introduction of a superfluous minus sign. Without loss of generality and to be consistent with the equations of \cite{MuskhelishviliNI:46a} and others, we here adopt the conventional definition.

This appendix examines the behavior of the complex velocity $w(\zeta)$ as the sharp edge is approached, $\zeta\rightarrow\zeta_o$. In general there are discontinuities in the strengths at $\zeta_o$ since this point represents the confluence of the sheets $S_1$, $S_2$ and $S_v$. As such, the sheet strength $\chi(\zeta)$ does not satisfy the H{\"o}lder condition \citep{MuskhelishviliNI:46a} exactly at $\zeta_o$ and so the Plemelj formulae for a corner point do not apply. However, each sheet strength may be said to satisfy the H{\"o}lder condition on its respective closed arc $S_1$, $S_2$ and $S_v$, in which case formulae are known for the behavior near the end points (see \textit{ibid}). 

The velocity induced by a given sheet at a point $\zeta$ near the corner point is:
\begin{eqnarray*}
w(\zeta)=\frac{1}{2\pi i}\int^{\zeta_o}\frac{\chi(\xi)}{\xi-\zeta}\md{\xi}&=&\frac{\chi(\zeta_o)}{2\pi i}\int^{\zeta_o}\frac{\md{\xi}}{\xi-\zeta}+\frac{1}{2\pi i}\int^{\zeta_o}\frac{\chi(\xi)-\chi(\zeta_o)}{\xi-\zeta}\md{\xi}\\
&=&\frac{\chi(\zeta_o)}{2\pi i}\log(\zeta-\zeta_o)+G(\zeta)
\end{eqnarray*}
where $G(\zeta)$ satisfies the H{\"o}lder condition near and at $\zeta_o$. The velocities of this form induced by the individual sheets are physically meaningless as $\zeta\rightarrow\zeta_o$, becoming logarithmically infinite, and the approaches from the left and right are not well defined. The singularity may be of a more general form if it is assumed that $\chi(\zeta)=\chi^{\ast}(\zeta)/(\zeta-\zeta_o)^\alpha$ near the corner where $\alpha\in\mathbb{C}$ is a complex constant. Although this may yield a finite $w(\zeta_o)$, we take the physical significance of the sheet strengths $\chi$ to imply that they too remain bounded, so that $\alpha=0$ and the singularity is logarithmic.

Therefore, we must consider the collective behavior of the total induced velocity. Let $\zeta_1$, $\zeta_2$ and $\zeta_v$ be positions on the sheets $S_1$, $S_2$ and $S_v$ such that $\epsilon=|\zeta_1-\zeta_o|=|\zeta_2-\zeta_o|=|\zeta_v-\zeta_o|$ are short equidistant lengths from the corner point $\zeta_o$. Then: 
\begin{eqnarray}
w(\zeta_o)=\frac{\chi_1+\chi_2-\chi_v}{2\pi i}\log\epsilon+G(\zeta_o) 
\end{eqnarray}
where $\chi_1$, $\chi_2$ and $\chi_v$ are each evaluated at $\zeta_o$. Hence, the condition of $\chi_v=\chi_1+\chi_2$ derived from mass and circulation conservation in \S\ref{sec:bint} ensures that this logarithmic singularity in the velocity is removed.

Now, the complex potential $\Phi(z)$ is determined by the same Cauchy-type integral as the conjugate velocity $w(z)$, so that $\Phi$ has an analogous logarithmic singularity. This is the source of the inverse square-root singularities in the velocity as mentioned by \cite{JonesMA:03a}. More specifically since $w=\md{\Phi}/\md{z}$, then:
\begin{equation}
\der{}{\zeta}\log|\zeta-\zeta_o|\sim\frac{1}{|\zeta-\zeta_o|}=\frac{1}{\sqrt{(x-x_o)^2+(y-y_o)^2}}.
\end{equation}
where recall that $\epsilon=|\zeta-\zeta_o|$. However, these singularities are also removed by the conditions imposed on the $\chi_j$. This is due to the relationship between the doublet sheet strength and the vortex-entrainment sheet strength as $\chi=\partial{\lambda}/\partial{\zeta}$. In fact, this is what allows the velocity to be written as:
\begin{equation*}
w(z)=\der{\Phi}{z}=\frac{1}{2\pi i}\int_{S}\frac{\chi(\zeta)\md{\zeta}}{z-\zeta}=\frac{1}{2\pi i}\int_{S}\frac{(\partial\lambda/\partial s)\md{s}}{z-\zeta(s)}.
\end{equation*}


\bibliographystyle{plain}


\begin{thebibliography}{46}
\expandafter\ifx\csname natexlab\endcsname\relax\def\natexlab#1{#1}\fi
\def\au#1{#1} \def\ed#1{#1} \def\yr#1{#1}\def\at#1{#1}\def\jt#1{\textit{#1}}
  \def\bt#1{#1}\def\bvol#1{\textbf{#1}} \def\vol#1{#1} \def\pg#1{#1}
  \def\publ#1{#1}\def\arxiv#1{#1}\def\org#1{#1}\def\st#1{\textit{#1}}

\bibitem[Alben(2008)]{AlbenS:08a}
{\sc \au{Alben, S.}} \yr{2008}  \at{Optimal flexibility of a flapping appendage
  in an inviscid fluid}.  \jt{Journal of Fluid Mechanics}  \bvol{614},
  \pg{355}.

\bibitem[Alben(2009)]{AlbenS:09a}
{\sc \au{Alben, S.}} \yr{2009}  \at{Simulating the dynamics of flexible bodies
  and vortex sheets}.  \jt{Journal of Computational Physics}  \bvol{228}~(7),
  \pg{2587--2603}.

\bibitem[Aris(1962)]{ArisR:62a}
{\sc \au{Aris, R.}} \yr{1962} {\em Vectors, Tensors and the Basic Equations of
  Fluid Mechanics\/}.  \publ{Englewood Cliffs, NJ, USA: Prentice-Hall}.

\bibitem[Baker {\em et~al.\/}(1982)Baker, Meiron \& Orszag]{Orszag:82a}
{\sc \au{Baker, G.~R.}, \au{Meiron, D.~I.} \& \au{Orszag, S.~A.}} \yr{1982}
  \at{Generalized vortex methods for free-surface flow problems}.  \jt{Journal
  of Fluid Mechanics}  \bvol{123},  \pg{477--501}.

\bibitem[Basu \& Hancock(1978)]{BasuBC:77a}
{\sc \au{Basu, B.~C.} \& \au{Hancock, G.~J.}} \yr{1978}  \at{The unsteady
  motion a two-dimensional aerofoil in incompressible inviscid flow}.
  \jt{Journal of Fluid Mechanics}  \bvol{87},  \pg{159--178}.

\bibitem[Batchelor(1967)]{Batchelor:67a}
{\sc \au{Batchelor, G.~K.}} \yr{1967} {\em An Introduction to Fluid
  Dynamics\/}.  \publ{Cambridge, UK: Cambridge University Press}.

\bibitem[Bhatia {\em et~al.\/}(2013)Bhatia, Norgard, Pascucci \&
  Bremer]{NorgardG:13a}
{\sc \au{Bhatia, H.}, \au{Norgard, G.}, \au{Pascucci, V.} \& \au{Bremer,
  P.-T.}} \yr{2013}  \at{The {Helmholtz-Hodge} decomposition - {A} survey}.
  \jt{{IEEE} Transactions of visualization and computer graphicsical
  Engineering Science}  \bvol{19}~(8),  \pg{1386--1404}.

\bibitem[Clements(1973)]{ClementsRR:73a}
{\sc \au{Clements, R.~R.}} \yr{1973}  \at{An inviscid model of two-dimensional
  vortex shedding}.  \jt{Journal of Fluid Mechanics}  \bvol{57}~(2),
  \pg{321--336}.

\bibitem[Cortelezzi \& Leonard(1993)]{LeonardA:93a}
{\sc \au{Cortelezzi, L.} \& \au{Leonard, A.}} \yr{1993}  \at{Point vortex model
  of the unsteady separated flow past a semi-infinite plate with transverse
  motion}.  \jt{Fluid Dynamics Research}  \bvol{11}~(6),  \pg{263--295}.

\bibitem[DeVoria \& Mohseni(2018)]{Mohseni:18l}
{\sc \au{DeVoria, A.~C.} \& \au{Mohseni, K.}} \yr{2018}  \at{Vortex sheet
  roll-up revisited}.  \jt{Journal of Fluid Mechanics}  \bvol{855},
  \pg{299--321}.

\bibitem[{DeVoria} \& Ringuette(2013)]{DeVoriaAC:13a}
{\sc \au{{DeVoria}, A.~C.} \& \au{Ringuette, M.~J.}} \yr{2013}  \at{On the flow
  generated on the leeward face of a rotating flat plate}.  \jt{Experiments in
  Fluids}  \bvol{54}~(4),  \pg{1--14}.

\bibitem[Giesing(1969)]{GiesingJP:69a}
{\sc \au{Giesing, J.~P.}} \yr{1969}  \at{Vorticity and {Kutta} condition for
  unsteady multienergy flows}.  \jt{{ASME} Journal of Applied Mechanics}
  \bvol{36}~(3),  \pg{608--613}.

\bibitem[Greenberg(1998)]{GreenbergM:98a}
{\sc \au{Greenberg, M.~D.}} \yr{1998} {\em Advanced Engineering Mathematics\/},
  2nd edn.  \publ{New Jersey: Prentice Hall}.

\bibitem[Haroldsen \& Meiron(1998)]{MeironD:98a}
{\sc \au{Haroldsen, D.~J.} \& \au{Meiron, D.~I.}} \yr{1998}  \at{Numerical
  calculation of three-dimensional interfacial potential flows using the point
  vortex method}.  \jt{{SIAM} Journal of Scientific Computing}  \bvol{20}~(2),
  \pg{648--683}.

\bibitem[Howe(2007)]{HoweMS:07a}
{\sc \au{Howe, M.~S.}} \yr{2007} {\em Hydrodynamics and sound\/}.
  \publ{Cambridge, UK: Cambridge University Press}.

\bibitem[Jackson(1998)]{JacksonJD:98a}
{\sc \au{Jackson, J.~D.}} \yr{1998} {\em Classical Electrodynamics\/}, 3rd edn.
   \publ{New York City, NY, USA: Wiley}.

\bibitem[Jones(2003)]{JonesMA:03a}
{\sc \au{Jones, M.~A.}} \yr{2003}  \at{The separated flow of an inviscid fluid
  around a moving flat plate}.  \jt{Journal of Fluid Mechanics}  \bvol{496},
  \pg{405--441}.

\bibitem[Katz(1981)]{KatzJ:81a}
{\sc \au{Katz, J.}} \yr{1981}  \at{A discrete vortex method for the non-steady
  separated flow over an airfoil}.  \jt{Journal of Fluid Mechanics}
  \bvol{102},  \pg{315--328}.

\bibitem[Kellogg(1929)]{KelloggO:29a}
{\sc \au{Kellogg, O.~D.}} \yr{1929} {\em Foundations of Potential Theory\/},
  1st edn.  \publ{Berlin: Springer}.

\bibitem[Koochesfahani(1989)]{Koochesfahani:89a}
{\sc \au{Koochesfahani, M.}} \yr{1989}  \at{Vortical patterns in the wake of an
  oscillating airfoil}.  \jt{{AIAA} Journal}  \bvol{27}~(9),  \pg{1200--1205}.

\bibitem[Lamb(1945)]{Lamb:45a}
{\sc \au{Lamb, H.}} \yr{1945} {\em Hydrodynamics\/}.  \publ{Mineola, NY, USA:
  Dover}.

\bibitem[Leonard(1980)]{Leonard:80a}
{\sc \au{Leonard, A.}} \yr{1980}  \at{Vortex methods for flow simulation}.
  \jt{Journal of Computational Physics}  \bvol{37}~(3),  \pg{289--335}.

\bibitem[Lighthill(1963)]{LighthillMJ:63a}
{\sc \au{Lighthill, M.J.}} \yr{1963}  \at{Introduction: Boundary layer theory}.
   \jt{In: Laminar Boundary Theory}  \pg{pp. 46--113}.

\bibitem[Liu {\em et~al.\/}(2015)Liu, Zhu \& Wu]{WuJZ:15a}
{\sc \au{Liu, L.~Q.}, \au{Zhu, J.~Y.} \& \au{Wu, J.~Z.}} \yr{2015}  \at{Lift
  and drag in two-dimensional steady viscous and compressible flow}.
  \jt{Journal of Fluid Mechanics}  \bvol{784},  \pg{304--341}.

\bibitem[Maskell(1971)]{MaskellEC:71a}
{\sc \au{Maskell, E.~C.}} \yr{1971}  \at{On the {Kutta-Joukowski} condition in
  two-dimensional unsteady flow}.  \jt{Unpublished note, Royal Aircraft
  Establishment, Farnborough, England} .

\bibitem[Michelin \& {Llewellyn Smith}(2009)]{LlewellynSmith:09a}
{\sc \au{Michelin, S.} \& \au{{Llewellyn Smith}, S.~G.}} \yr{2009}  \at{An
  unsteady point vortex method for coupled fluid-solid problems}.
  \jt{Theoretical and Computational Fluid Dynamics}  \bvol{23}~(2),
  \pg{127--153}.

\bibitem[Muskhelishvili(1946)]{MuskhelishviliNI:46a}
{\sc \au{Muskhelishvili, N.~I.}} \yr{1946} {\em Singular Integral Equations\/},
  1st edn.  \publ{Moscow: P. Noordhoof Ltd.}

\bibitem[Nitsche \& Krasny(1994)]{Nitsche:94a}
{\sc \au{Nitsche, M.} \& \au{Krasny, R.}} \yr{1994}  \at{A numerical study of
  vortex ring formation at the edge of a circular tube}.  \jt{Journal of Fluid
  Mechanics}  \bvol{276},  \pg{139--161}.

\bibitem[Phillips(1959)]{PhillipsHB:59a}
{\sc \au{Phillips, H.~B.}} \yr{1959} {\em Vector Analysis\/}, 1st edn.
  \publ{John Wiley \& Sons Inc.}, 18th printing.

\bibitem[Poling \& Telionis(1986)]{PolingDR:86a}
{\sc \au{Poling, D.~R.} \& \au{Telionis, D.~P.}} \yr{1986}  \at{The response of
  airfoils to periodic disturbances: {The} unsteady {Kutta} condition}.
  \jt{{AIAA} Journal}  \bvol{24}~(2),  \pg{193--199}.

\bibitem[Pozrikidis(2000)]{Pozrikidis:00a}
{\sc \au{Pozrikidis, C.}} \yr{2000}  \at{Theoretical and computation aspects of
  the self-induced motion of three-dimensional vortex sheets}.  \jt{Journal of
  Fluid Mechanics}  \bvol{425},  \pg{335--366}.

\bibitem[Pullin(1978)]{Pullin:78a}
{\sc \au{Pullin, D.~I.}} \yr{1978}  \at{The large-scale structure of unsteady
  self-similar rolled-up vortex sheets}.  \jt{Journal of Fluid Mechanics}
  \bvol{88}~(3),  \pg{401--430}.

\bibitem[Rosenhead(1931)]{RosenheadL:31a}
{\sc \au{Rosenhead, L.}} \yr{1931}  \at{The formation of vortices from a
  surface of discontinuity}.  \jt{Proceedings of the Royal Society A:
  Mathematical, Physical and Engineering Sciences}  \bvol{134}~(823),
  \pg{170--192}.

\bibitem[Rott(1956)]{RottN:56a}
{\sc \au{Rott, N.}} \yr{1956}  \at{Diffraction of a weak shock with vortex
  generation}.  \jt{Journal of Fluid Mechanics}  \bvol{1},  \pg{111--128}.

\bibitem[Saffman(1992)]{Saffman:92a}
{\sc \au{Saffman, P.~G.}} \yr{1992} {\em Vortex Dynamics\/}.  \publ{Cambridge,
  UK: Cambridge University Press}.

\bibitem[Sarpkaya(1975)]{SarpkayaT:75a}
{\sc \au{Sarpkaya, T.}} \yr{1975}  \at{An inviscid model of two-dimensional
  vortex shedding for transient and asymptotically steady separated flow over
  an inclined plate}.  \jt{Journal of Fluid Mechanics}  \bvol{68},
  \pg{109--128}.

\bibitem[Schlichting(1955)]{Schlichting:55a}
{\sc \au{Schlichting, H.}} \yr{1955} {\em Boundary Layer Theory\/}, 2nd edn.
  \publ{New York: McGraw-Hill}.

\bibitem[Scriven(1960)]{ScrivenLE:60a}
{\sc \au{Scriven, L.~E.}} \yr{1960}  \at{Dynamics of a fluid interface equation
  of motion for newtonian surface fluids}.  \jt{Chemical Engineering Science}
  \bvol{12}~(2),  \pg{98--108}.

\bibitem[Sears(1956)]{SearsWR:56a}
{\sc \au{Sears, W.~R.}} \yr{1956}  \at{Some recent developments in airfoil
  theory}.  \jt{Journal of the Aeronautical Sciences}  \bvol{23}~(5),
  \pg{490--499}.

\bibitem[Slattery {\em et~al.\/}(2007)Slattery, Sagis \& Oh]{SlatteryJC:07a}
{\sc \au{Slattery, J.~C.}, \au{Sagis, L.} \& \au{Oh, E.~S.}} \yr{2007} {\em
  Interfacial Transport Phenomena\/}, 2nd edn.  \publ{New York, NY, USA:
  Springer, New York}.

\bibitem[Stakgold(1968)]{StakgoldI:68a}
{\sc \au{Stakgold, I.}} \yr{1968} {\em Boundary Value Problems of Mathematical
  Physics: Volume 2\/}, 1st edn.  \publ{Society for Industrial and Applied
  Mathematics}.

\bibitem[Wang \& Eldredge(2013)]{EldredgeJD:13a}
{\sc \au{Wang, C.} \& \au{Eldredge, J.~D.}} \yr{2013}  \at{Low-order
  phenomenological modeling of leading-edge vortex formation}.  \jt{Theoretical
  and Computational Fluid Dynamics}  \bvol{27}~(5),  \pg{577--598}.

\bibitem[Westwater(1935)]{WestwaterF:35a}
{\sc \au{Westwater, F.~L.}} \yr{1935}  \bt{The rolling up of a surface of
  discontinuity behind an aerofoil of finite span}. Report R\&M 1692.
  \org{Aeronautical Research Council}.

\bibitem[Wu {\em et~al.\/}(2006)Wu, Ma \& Zhou]{WuJZ:06a}
{\sc \au{Wu, J-Z.}, \au{Ma, H-Y.} \& \au{Zhou, M-D.}} \yr{2006} {\em Vorticity
  and vortex dynamics\/}.  \publ{Springer}.

\bibitem[Xia \& Mohseni(2013)]{Mohseni:13ag}
{\sc \au{Xia, X.} \& \au{Mohseni, K.}} \yr{2013}  \at{Lift evaluation of a
  two-dimensional pitching flat plate}.  \jt{Physics of Fluids}  \bvol{25}~(9),
   \pg{091901(1--26)}.

\bibitem[Xia \& Mohseni(2017)]{Mohseni:17m}
{\sc \au{Xia, X.} \& \au{Mohseni, K.}} \yr{2017}  \at{Unsteady aerodynamics and
  vortex-sheet formation of a two-dimensional airfoil}.  \jt{Journal of Fluid
  Mechanics}  \bvol{830},  \pg{439--478}.

\end{thebibliography}

\end{document}